\documentstyle[12pt]{article}
\begin{document}
\newcommand{\preprint}[1]{
\begin{table}[t]
\begin{flushright}
\begin{large}{#1}\end{large}
\end{flushright}
\end{table}}
\baselineskip 18pt
\preprint{TAUP-2280-95}
\preprint{IASSNS-HEP-96/71}
\newcommand{\bce}{\begin{center}}
\newcommand{\ece}{\end{center}}

\title{A Chiral Spin Theory in the Framework of an Invariant Evolution
Parameter Formalism}
\author{B.Sarel\\ School of Physics and Astronomy\\ Raymond and
Beveryly Sacklar Faculty of Exact Sciences\\ Tel Aviv University, 69978
Ramat Aviv, Israel \\and \\L.P.Horwitz\thanks{On sabbatical leave from
School of Physics and Astronomy, Raymond and
Beverly Sackler Faculty of Exact Sciences, Tel Aviv University, Ramat
Aviv, Israel and Department of Physics, Bar Ilan University, Ramat
Gan, Israel}\\School of Natural Sciences, Institute for Advanced
Study\\ Princeton, N.J. 08540}
\maketitle

\newcommand{\schstu}{Schr\"{o}dinger-Stueckelberg\/ }

\begin{abstract}
We present a formulation for the construction of first order equations which
describe particles with spin, in the context of a manifestly covariant
relativistic theory governed by an invariant evolution parameter; one obtains
a consistent quantized formalism dealing with off-shell particles
with spin. Our basic requirement is that the second order equation in the
theory is of the \schstu type, which exhibits features of both the
Klein-Gordon and Schr\"{o}dinger equations.
This requirement restricts the structure of the first order equation, in
particular, to a chiral form.
One thus obtains, in a natural way, a theory of chiral form for massive
particles, which may contain both left and right chiralities, or just one of
them.
We observe that by iterating the first order system,
we are able to obtain second order forms containing the transverse and
longitudinal momentum relative to a time-like vector $t_{\mu}t^{\mu}=-1$ used
to maintain covariance of the theory. This time-like vector coincides with
the one used by Horwitz, Piron, and Reuse to obtain an invariant
positive definite space-time scalar product, which permits the construction
of an induced representation for states of a particle with spin.
We discuss the currents and continuity equations,
and show that these equations of motion and their currents are closely related
to the spin and convection parts of the Gordon decomposition of the Dirac
current.
The transverse and longitudinal aspects of the particle are complementary,
and can be treated in a unified manner using a tensor product Hilbert space.
Introducing the
electromagnetic field we find an equation which gives rise to the correct
gyromagnetic ratio, and is fully Hermitian under the proposed scalar product.
Finally, we
show that the original structure of Dirac's equation and its solutions is
obtained in the highly constrained limit in which $p_{\mu}$ is proportional to
$t_{\mu}$ on mass shell.
The chiral nature of the theory is apparent. We
define the discrete symmetries of the theory, and find that
they are represented by states which are pure left or right handed.
\end{abstract}
\newpage

\newcommand{\myx}{\mbox{$\times$}}

\newcommand{\spinhalf}{\mbox{spin $\frac{1}{2}$ }}
\newcommand{\dtau}{\mbox{$\partial_{\tau}$}}
\newcommand{\idtau}{\mbox{$i\dtau$}}
\newcommand{\splus}{\mbox{$s_{+}$}}
\newcommand{\sminus}{\mbox{$s_{-}$}}
\newcommand{\spn}{\mbox{$\splus\sminus$}}
\newcommand{\nplus}{\mbox{$N_{+}$}}
\newcommand{\nminus}{\mbox{$N_{-}$}}
\newcommand{\pplus}{\mbox{$({\bf 1}+\gamma^{5})$}}
\newcommand{\pminus}{\mbox{$({\bf 1}-\gamma^{5})$}}
\newcommand{\gamadt}{\mbox{$(\gamma\cdot t)$}}
\newcommand{\gdt}{\mbox{$(\gamma\cdot t)$}}
\newcommand{\gds}{\mbox{$(\gamma\cdot s)$}}
\newcommand{\egdt}{\mbox{$(\gamma^{\mu}\cdot t_{\mu})$}}
\newcommand{\sigpta}{\mbox{$\sigma^{\mu\nu}P_{\mu}t_{\nu}$}}
\newcommand{\enplus}{\mbox{$\gdt\pplus$}}
\newcommand{\enminus}{\mbox{$\gdt\pminus$}}

\newcommand{\omegamunu}{\mbox{$\omega_{\mu\nu}$}}
\newcommand{\sigmunu}{\mbox{$\sigma^{\mu\nu}$}}
\newcommand{\sigrolam}{\mbox{$\sigma^{\rho\lambda}$}}
\newcommand{\sigmuro}{\mbox{$\sigma^{\mu\rho}$}}
\newcommand{\sigmulam}{\mbox{$\sigma^{\mu\lambda}$}}
\newcommand{\signuro}{\mbox{$\sigma^{\nu\rho}$}}
\newcommand{\signulam}{\mbox{$\sigma^{\nu\lambda}$}}

\newcommand{\sigmzi}{\mbox{$\sigma^{0i}$}}
\newcommand{\sigmiz}{\mbox{\$sigma^{i0}$}}
\newcommand{\sigmij}{\mbox{$\sigma^{ij}$}}
\newcommand{\gamaf}{\mbox{$\gamma^{5}$}}
\newcommand{\gamamu}{\mbox{$\gamma^{\mu}$}}
\newcommand{\gamanu}{\mbox{$\gamma^{\nu}$}}
\newcommand{\gamaro}{\mbox{$\gamma^{\rho}$}}
\newcommand{\gamalam}{\mbox{$\gamma^{\lambda}$}}
\newcommand{\dmu}{\mbox{$\partial_{\mu}$}}
\newcommand{\dnu}{\mbox{$\partial_{\nu}$}}
\newcommand{\sigmapv}{\mbox{$\sigma Pv$}}
\newcommand{\sigmapt}{\mbox{$(\sigma Pt)$}}
\newcommand{\sigmaptp}{\mbox{$\sigma Pt^{'}$}}
\newcommand{\sigmavv}{\mbox{$\sigma v^{1}v^{2}$}}
\newcommand{\sigmats}{\mbox{$\sigma ts$}}

\newcommand{\gmunu}{\mbox{$g^{\mu\nu}$}}
\newcommand{\grolam}{\mbox{$g^{\rho\lambda}$}}
\newcommand{\gmuro}{\mbox{$g^{\mu\rho}$}}
\newcommand{\gmulam}{\mbox{$g^{\mu\lambda}$}}
\newcommand{\gnuro}{\mbox{$g^{\nu\rho}$}}
\newcommand{\gnulam}{\mbox{$g^{\nu\lambda}$}}

\newcommand{\pdttre}{\mbox{$({\bf P\cdot t^{\prime}})$}}
\newcommand{\pdtfour}{\mbox{$(P\cdot t^{\prime})$}}

\newcommand{\gamaz}{\mbox{$\gamma^{0}$}}
\newcommand{\gamaone}{\mbox{$\gamma^{1}$}}
\newcommand{\gamatwo}{\mbox{$\gamma^{2}$}}
\newcommand{\gamatre}{\mbox{$\gamma^{3}$}}

\newcommand{\Gamai}{\mbox{$\Gamma_{i}$}}
\newcommand{\Gamaj}{\mbox{$\Gamma_{j}$}}
\newcommand{\Gamak}{\mbox{$\Gamma_{k}$}}
\newcommand{\Gamal}{\mbox{$\Gamma_{l}$}}
\newcommand{\Pip}{\mbox{$P_{i\mbox{+}}$}}
\newcommand{\Pin}{\mbox{$P_{i\mbox{-}}$}}
\newcommand{\Pjp}{\mbox{$P_{j\mbox{+}}$}}
\newcommand{\Pjn}{\mbox{$P_{j\mbox{-}}$}}
\newcommand{\Ti}{\mbox{$T_{i}$}}
\newcommand{\Tj}{\mbox{$T_{j}$}}

\newcommand{\psip}{\mbox{$\psi_{\perp}$}}
\newcommand{\psil}{\mbox{$\psi_{\parallel}$}}
\newcommand{\varphip}{\mbox{$\varphi_{\perp}$}}
\newcommand{\varphil}{\mbox{$\varphi_{\parallel}$}}

\newcommand{\psidag}{\mbox{$\psi^{\dagger}$}}
\newcommand{\psibar}{\mbox{$\bar{\psi}$}}
\newcommand{\phidag}{\mbox{$\phi^{\dagger}$}}
\newcommand{\phibar}{\mbox{$\bar{\phi}$}}

\newcommand{\psidagp}{\mbox{$\psi_{\perp}^{\dagger}$}}
\newcommand{\psibarp}{\mbox{$\bar{\psi}_{\perp}$}}
\newcommand{\phidagp}{\mbox{$\phi_{\perp}^{\dagger}$}}
\newcommand{\phibarp}{\mbox{$\bar{\phi}_{\perp}$}}

\newcommand{\psidagl}{\mbox{$\psi^_{\parallel}{\dagger}$}}
\newcommand{\psibarl}{\mbox{$\bar{\psi}_{\parallel}$}}
\newcommand{\phidagl}{\mbox{$\phi_{\parallel}^{\dagger}$}}
\newcommand{\phibarl}{\mbox{$\bar{\phi}_{\parallel}$}}

\newcommand{\PsiL}{\mbox{$\Psi_{L}$}}
\newcommand{\PsiR}{\mbox{$\Psi_{R}$}}

\newcommand{\psiL}{\mbox{$\psi_{L}$}}
\newcommand{\psiR}{\mbox{$\psi_{R}$}}
\newcommand{\psiLbar}{\mbox{$\bar{\psi}_{L}$}}
\newcommand{\psiRbar}{\mbox{$\bar{\psi}_{R}$}}

\newcommand{\psiLp}{\mbox{$\psi_{\perp L}$}}
\newcommand{\psiRp}{\mbox{$\psi_{\perp R}$}}
\newcommand{\psiLbarp}{\mbox{$\bar{\psi}_{\perp L}$}}
\newcommand{\psiRbarp}{\mbox{$\bar{\psi}_{\perp R}$}}

\newcommand{\psiLl}{\mbox{$\psi_{\parallel L}$}}
\newcommand{\psiRl}{\mbox{$\psi_{\parallel R}$}}
\newcommand{\psiLbarl}{\mbox{$\bar{\psi}_{\parallel L}$}}
\newcommand{\psiRbarl}{\mbox{$\bar{\psi}_{\parallel R}$}}

\newcommand{\phip}{\mbox{$\phi_{\perp}$}}
\newcommand{\phil}{\mbox{$\phi_{\parallel}$}}
\newcommand{\chip}{\mbox{$\chi_{\perp}$}}
\newcommand{\chil}{\mbox{$\chi_{\parallel}$}}

\newcommand{\varphidag}{\mbox{$\varphi^{\dagger}$}}
\newcommand{\varphibar}{\mbox{$\bar{\varphi}$}}
\newcommand{\varphiL}{\mbox{$\varphi_{L}$}}
\newcommand{\varphiR}{\mbox{$\varphi_{R}$}}
\newcommand{\varphiLbar}{\mbox{$\bar{\varphi}_{L}$}}
\newcommand{\varphiRbar}{\mbox{$\bar{\varphi}_{R}$}}

\newcommand{\varphidagp}{\mbox{$\varphi_{\perp}^{\dagger}$}}
\newcommand{\varphibarp}{\mbox{$\bar{\varphi}_{\perp}$}}
\newcommand{\varphiLp}{\mbox{$\varphi_{\perp L}$}}
\newcommand{\varphiRp}{\mbox{$\varphi_{\perp R}$}}
\newcommand{\varphiLbarp}{\mbox{$\bar{\varphi}_{\perp L}$}}
\newcommand{\varphiRbarp}{\mbox{$\bar{\varphi}_{\perp R}$}}

\newcommand{\varphidagl}{\mbox{$\varphi_{\parallel}^{\dagger}$}}
\newcommand{\varphibarl}{\mbox{$\bar{\varphi}_{\parallel}$}}
\newcommand{\varphiLl}{\mbox{$\varphi_{\parallel L}$}}
\newcommand{\varphiRl}{\mbox{$\varphi_{\parallel R}$}}
\newcommand{\varphiLbarl}{\mbox{$\bar{\varphi}_{\parallel L}$}}
\newcommand{\varphiRbarl}{\mbox{$\bar{\varphi}_{\parallel R}$}}

\newcommand{\antiCLp}{\mbox{$\tilde{C}_{l+}$}}
\newcommand{\antiCLn}{\mbox{$\tilde{C}_{l-}$}}
\newcommand{\antiCnn}{\mbox{$\tilde{C}_{nn}$}}
\newcommand{\CLp}{\mbox{$C_{l+}$}}
\newcommand{\CLn}{\mbox{$C_{l-}$}}
\newcommand{\Cnn}{\mbox{$C_{nn}$}}

\newcommand{\sigpit}{\mbox{$(\sigma\Pi t)$}}
\newcommand{\tsigpit}{\mbox{$(\sigma\tilde{\Pi}t)$}}
\newcommand{\esigpit}{\mbox{$(\sigmunu P_{\mu}t_{\nu})$}}
\newcommand{\atau}{\mbox{$a_{\tau}$}}
\newcommand{\tpsik}{\mbox{$t^{\prime}$}}
\newcommand{\ppsik}{\mbox{$p^{\prime}$}}
\newcommand{\Ppsik}{\mbox{$P^{\prime}$}}

\newcommand{\jmuLp}{\mbox{$j^{\mu}_{\perp L}$}}
\newcommand{\jmuRp}{\mbox{$j^{\mu}_{\perp R}$}}
\newcommand{\jmuLl}{\mbox{$j^{\mu}_{\parallel L}$}}
\newcommand{\jmuRl}{\mbox{$j^{\mu}_{\parallel R}$}}
\newcommand{\sigmaptop}{\mbox{$(\sigma\!\! \stackrel{\leftarrow}{P}\! t)$}}
\newcommand{\LRdmu}{\mbox{$\stackrel{\leftrightarrow}{\partial^{\mu}}$}}
\newcommand{\LRdmud}{\mbox{$\stackrel{\leftrightarrow}{\partial_{\mu}}$}}
\newcommand{\LRdmudt}{\mbox{$\stackrel{\leftrightarrow}
{\partial_{\perp\mu}}$}}
\newcommand{\LRdmudl}{\mbox{$\stackrel{\leftrightarrow}
{\partial_{\parallel\mu}}$}}

\newcommand{\ssigpt}{\mbox{$(\sigma pt)$}}
\newcommand{\hfive}{\mbox{$i\gdt\gamaf$}}
\newcommand{\sigptgdt}{\mbox{$\ssigpt\gdt$}}
\newcommand{\helicity}{\mbox{$\frac{i}{\mid p_{\perp}\mid}\ssigpt\gamaf$}}
\newcommand{\hpt}{\mbox{$\frac{1}{\mid p_{\perp}\mid}\ssigpt\gdt$}}
\newcommand{\htt}{\mbox{$\gdt$}}
\newcommand{\hptf}{\mbox{$\htt\helicity$}}

\newcommand{\Pitild}{\mbox{$\stackrel{\leftarrow}{\tilde{\Pi}}$}}
\newcommand{\Pitildmu}{\mbox{$\stackrel{\!\!\!\!\leftarrow}
{\tilde{\Pi}^{\mu}}$}}
\newcommand{\Pitildt}{\mbox{$\stackrel{\!\!\!\!\leftarrow}
{\tilde{\Pi}^{2}}$}}
\newcommand{\Pinorm}{\mbox{$\stackrel{\rightarrow}{\Pi}$}}
\newcommand{\Pinormt}{\mbox{$\stackrel{\!\!\!\!\rightarrow}{\Pi^{2}}$}}
\newcommand{\Piboth}{\mbox{$\stackrel{\leftrightarrow}{\Pi}$}}
\newcommand{\Pinormmu}{\mbox{$\stackrel{\!\!\!\!\rightarrow}{\Pi^{\mu}}$}}
\newcommand{\Pibothmu}{\mbox{$\stackrel{\!\!\!\!\leftrightarrow}
{\Pi^{\mu}}$}}
\newcommand{\Pibothmup}{\mbox{$\stackrel{\!\!\!\!\leftrightarrow}
{\Pi_{\perp}^{\mu}}$}}
\newcommand{\Pinormup}{\mbox{$\stackrel{\!\!\!\!\rightarrow}
{\Pi_{\perp}^{\mu}}$}}
\newcommand{\Pinormul}{\mbox{$\stackrel{\!\!\!\!\rightarrow}
{\Pi_{\parallel}^{\mu}}$}}
\newcommand{\Pitildmup}{\mbox{$\stackrel{\!\!\!\!\leftarrow}
{\tilde{\Pi}_{\perp}^{\mu}}$}}
\newcommand{\Pitildmul}{\mbox{$\stackrel{\!\!\!\!\leftarrow}
{\tilde{\Pi}_{\parallel}^{\mu}}$}}
\newcommand{\Pibothmul}{\mbox{$\stackrel{\!\!\!\!\leftrightarrow}
{\Pi_{\parallel}^{\mu}}$}}


\section*{I. Introduction}
\label{sec:intro}

The Dirac theory of the \spinhalf particles, like the Klein-Gordon description
of particles without spin, makes use of wave functions which are not
localized,
as pointed out by Newton and Wigner \cite{NewWig}. The equations of motion,
although useful for describing the properties of quantum fields, are not
adequate for the construction of effective one-particle quantum theories. It
has been found that the use of an invariant parameter to describe the
evolution
of states which are off-mass-shell provides a general framework in which
this problem is solved \cite{HP73,AH85}.

First order equations for particles with \spinhalf, for which the
evolution of the entire system is governed by an invariant parameter, have
been
searched for and studied by many over the years. Their importance lies in
describing the behavior of the spin degrees of freedom of fermionic particles
in such theories. We can trace the formulation of covariant
quantum theories with an invariant evolution parameter from Fock
\cite{Fock37},
through Stueckelberg \cite{Stu4142}, Nambu \cite{Nambu50}, Schwinger
\cite{Schw50}, and Feynman \cite{Fey5051}, up to more recent work done by
Cooke \cite{Cooke68}, Horwitz and Piron \cite{HP73}, and Fanchi \cite{Fan86}.

Second order equations for \spinhalf particles have been found and studied by
Fock \cite{Fock37}, Feynman \cite{Fey5051}, Horwitz {\sl et al} \cite{HPR75},
and Reuse \cite{Reuse78}.
A number of first order equations for \spinhalf particles have been proposed
by Nambu \cite{Nambu50}, Feynman \cite{Fey5051}, Kubo \cite{Kubo85}, and
Davidon \cite{DAV55}. The Dirac equation \cite{Dirac28} does not contain an
invariant
evolution parameter, and applies to a three-dimensional measure space.
For a summary of the subject and
an extensive list of references see Fanchi \cite{Fan93}. Some of the first
order equations were introduced in an {\sl ad hoc} manner, and some of the
formulations were
incompatible with the postulated second order equation for the evolution of a
free particle in the corresponding theories.
In some cases they led to free evolution equations
which are {\sl second order} in the invariant parameter,
admitting solutions which propagate forward and backward,
thus invalidating the interpretation of the invariant
parameter as an {\sl unidirectional} evolution parameter \cite{fey49,gravity},
a feature that is quite important for the interpretation of the theory.

In this paper we propose a first order equation of motion for a \spinhalf
par\-ticle, in the framework of the formalism developed by Horwitz and Piron
\cite{HP73}, which is consistent with the form of the second order evolution
equation for a free particle, the \schstu\/ equation
\begin{equation}
\idtau\psi=\frac{P^{\mu}P_{\mu}}{2M}\psi    \label{eq:sse}
\end{equation}
It seems that this kind of evolution kernel, proportional to $P^{\mu}P_{\mu}$,
is best suited to describe a covariant theory (see Fanchi \cite{Fan94}).
It leads
to the correct relation for the velocities of a classical particle and permits
separation of variables in the many body case \cite{HP73}.

We find that to achieve the goal of obtaining a first order equation for a
\spinhalf particle, it is necessary to introduce nilpotent
operators which result in a chiral theory even though we are not restricted
to massless fermions. This result may be relevant to the present situation
in weak interactions and neutrino physics, where one finds chiral fields
although it is not clear the neutrino mass is zero, and when
theories beyond the standard model are considered \cite{OT95,Moe95,Sciama}.
We shall restrict our study of interactions here, however, to the case of
the U(1) electromagnetic gauge.

The general outline of this work is as follows.
In Section II  we present a short summary of the basics of the
formalism we use. We then
establish our basic requirements of the equation, and the logic in deriving
it.
In Section III we define the basic structure of the equations
of motion, the form of a continuity equation, currents,
and probability density. The basic requirements force us to seek
nilpotent matrices as building blocks for the equation, so we investigate
all possible nilpotent $4\times 4$ matrices in Appendix \ref{app:all_potents},
(and state some useful facts about them),
and we state the Lorentz covariant appropriate nilpotent forms.
We observe that it is not possible to construct a single
\schstu type equation by iteration. We manage however, in Section
IV,
to construct two equations of motion, for transverse and longitudinal modes.
After doing that, we still have four possible forms for both versions,
and we show the relations among them. Two of them
vary by the relative signs of their chiral components, and the other two are
complementary in a sense that is described
later on. In Section V, we investigate the form
of the resulting currents and establish the validity of the form of the
probability
density for both versions. At this stage we observe the connection between
the transverse and longitudinal equations, and the spin and convection
currents, respectively, of the Gordon decomposition of the Dirac current.
In Section VI we define the tensor
product Hilbert space over the longitudinal and transverse modes, and show
that our formulation is fully compatible
with the one of Horwitz and Arshansky \cite{HA82}.
Next, in Section VII, we introduce a minimal coupling
for the gauge field, producing electromagnetic interaction. We discuss the
second
order form of the generator of motion, and show it coincides with the one
achieved by Horwitz and Arshansky \cite{HA82}.
It implies the correct gyromagnetic ratio, and is fully Hermitian under the
positive definite, invariant scalar product for the quantum mechanical Hilbert
space. In Section VIII, we show that we recover
the original Dirac equation and its solutions in the limit where $p_{\mu}$ is
constrained to be proportional to $t_{\mu}$ on mass shell.
The solutions of the equations
of motion are discussed in Appendix \ref{app:eqnsols}. We explore the
discrete symmetries
of the theory, in Appendix \ref{app:pct}, and define the generalized parity,
charge conjugation, and $\tau$ reversal transformations.
The states that are transformed one into the other
under these transformations are pure left or right handed. Finally, in
Appendix
\ref{app:diracmat} we make explicit the representation of the Dirac matrices
we used for calculations, which lends itself to the problem, the chiral
representation.

We use the notation
\begin{displaymath}
x^{\mu}\equiv (x^{0},x^{1},x^{2},x^{3})
\end{displaymath}
and we shall always use the symbol $x^{0}$ as the time component, and reserve
$t^{\mu}$ for the time-like vector. We use an opposite metric
relative to the Bjorken-Drell convention \cite{BD}
\begin{displaymath}
g_{00}= -1 = -g_{11}= -g_{22}= -g_{33}
\end{displaymath}
Furthermore, we shall use the notation
\begin{displaymath}
(\gamma\cdot v)\equiv(\gamma^{\mu} v_{\mu})
\end{displaymath}
where $v_{\mu}$ is a four-vector. We also use the uppercase $P_{\mu}$
to denote an operator, and lowercase $p_{\mu}$ for eigenvalues.


\section*{II. $\tau$-formalism}
\label{sec:tauform}

We refer to the invariant parameter as $\tau$, the invariant universal world
time, which describes the evolution of an event moving through space-time.
The notion of the need of some invariant evolution parameter to replace the
covariant time is not new, and we give some of the arguments as presented in
\cite{Hor92}.

Non-relativistic quantum mechanics uses the Newtonian universal
time to describe a state in terms of square integrable functions over
three-space at a specific time, which evolve according to Schr\"{o}dinger's
equation. But the Hilbert spaces associated with different times are distinct;
we cannot superpose wave functions at different times.
This situation is inconsistent with special relativity.
Viewed from some other frame the wave function becomes a function on different
times, thus losing its interpretation as a state.

The Klein-Gordon and Dirac equations have been able to resolve the problems
of covariance; they have a manifestly covariant form.
But the problem of constructing {\it localized}
states still remains. The inconsistency of the solutions of the Klein-Gordon
equation as an amplitude for a local probability density was shown by
Newton and Wigner \cite{NewWig}. They showed that the distribution
corresponding to a localized particle is an eigenfunction of the operator
\begin{equation}
X_{NW} = i(\frac{\partial}{\partial{\bf p}}-\frac{{\bf p}}{2E^{2}})
\end{equation}
and the wave function corresponding to a localized particle is spread out by
the order of a Compton wavelength.
They reached similar conclusions concerning the Dirac equation.
Hegerfeldt \cite{Heg}, has shown that a distribution at a specific time,
defined to
have compact support, or localized in some other sense, does not maintain its
localizability, and evolves out of the light cone, i.e. acausally.

Quantum field theories make the transformation laws of special relativity and
quantum mechanics consistent by assigning the spatial variables to the same
parametric
role as the time. The dynamical variables are operator valued fields which are
functions on this parametric space-time. But the one particle sector wave
functions of such theories describing the transition amplitudes between the
vacuum and one particle states, suffers from the same difficulties as
mentioned above.
Predictions of phenomena concerning local properties in space-time are
very difficult to formulate and interpret (e.g. interference phenomena),
although
spectral properties of non-local observables (energy, Lamb shift, anomalous
magnetic moment), can be computed and are in excellent agreement with
experiment.

The basic difficulty of developing a consistent theory which
incorporates the ideas of special relativity and quantum mechanics is related
to questions concerning the relationship between time and locality.
On one hand, space and time transform with the Lorentz group (and thus
this relativistic time has a geometrical interpretation), and on the other
hand, in a specific frame, it has been considered as a measure of evolution,
of change.
A way to resolve this ambiguity is to define
the state of a system in terms of a distribution of {\sl events} in space
{\sl and}
time, while their evolution is parameterized by the time indicated on an ideal
clock which is associated with
every inertial frame (see \cite{HAE88} for further discussion).
We call this parameter the
universal time $\tau$, and it can be identified with Newton's time.

Stueckelberg \cite{Stu4142}, Horwitz and Piron \cite{HP73}, and others
\cite{Schw50,Fey5051,Fan86} developed an underlying formalism incorporating
the invariant parameter $\tau$ which enabled them to construct a consistent
manifestly covariant relativistic classical and quantum theory.
The Newton-Wigner \cite{HP73} as well as the Landau-Peierls \cite{H85}
problems have been understood in this framework. Two body problems, both for
bound states and scattering have been treated, and the Zeeman \cite{LH95}
effect
and selection rules \cite{LH94} for radiation worked out. The spinless theory,
in interaction with radiation (U(1) gauge field) has been second quantized
\cite{ShnerbH}.
We state briefly the main principles of the formalism.

The equations of motion of the classical theory, may be derived from the
Hamilton principle \cite{HP73}
\begin{equation}
\delta\;\int(p_{\mu}dq^{\mu}-K(p^{\mu},q^{\mu})d\tau)=0  \label{dayprinc}
\end{equation}
This principle is equivalent to the canonical equations
\begin{eqnarray}
\frac{dp_{\mu}}{d\tau} & = & -\frac{\partial K}{\partial q^{\mu}} \nonumber \\
\frac{dq^{\mu}}{d\tau} & = &  \;\;\frac{\partial K}{\partial p_{\mu}}
\end{eqnarray}
describing the motion of an ``event'' along its world line (trajectory).
For example, for a free event one takes
\begin{equation}
K_{0}=\frac{p^{\mu}p_{\mu}}{2M}=\frac{{\bf p}^{2}-E^{2}}{2M}
\end{equation}
where $M$ is a given property of the event, and sets the scale between $\tau$
and the quantities of motion. We then have
\begin{eqnarray}
\frac{d{\bf x}}{d\tau}=\frac{{\bf p}}{M} & ; & \frac{dt}{d\tau}=\frac{E}{M}
\end{eqnarray}
The proper time interval for the motion of a free event is defined by
\begin{equation}
ds^{2}=dt^{2}-d{\bf x}^{2}
\end{equation}
and satisfies
\begin{equation}
ds^{2}=\frac{m^{2}}{M^{2}}d\tau^{2}
\end{equation}
where $m^{2}=E^{2}-{\bf p}^{2}$ is a dynamical variable to be determined by
initial conditions and dynamics of the system. If initial conditions are
chosen
so that $m^{2}=M^{2}$, the ``on-shell'' condition, then the proper time
interval
and the universal world time $\tau$, coincide. It should be noted that the
theory
is not constrained to time-like motion, so tachyonic propagation of events is
possible.
However, this does not imply explicitly the existence of tachyonic particles
in laboratory measurements since one generally observes asymptotic states as
the initial and final outcome of collision experiments.
Due to the asymptotic conservation of the generator of free propagation
$\frac{P^{\mu}P_{\mu}}{2M}$, for example, in potential scattering, states that
are initially in time-like motion are time-like in the asymptotic final
state as well. The structure of the theory, however, does not exclude
tachyons {\sl a priori}, classically or quantum mechanically.

In the quantum domain, the states of the system for a given $\tau$ are
described
in the Hilbert space $L^{2}(R^{4},d^{3}xdt)$, the space of square integrable
functions of four variables given for the spinless case as
$(\psi,\chi)=\int\psi^{\ast}\chi d^{4}x$.
We do not define the scalar product for \spinhalf particles at this stage;
instead we shall show that we obtain the scalar product suggested by
Horwitz and Arshansky \cite{HA82}, and Arensburg and Horwitz \cite{HAR90},
from considerations of the equation of continuity.

The observables of spacetime coordinates and momenta satisfy the commutation
relations
\begin{equation}
i[P^{\mu},Q^{\nu}]=g^{\mu\nu}{\bf 1}  \label{eq:canonvar}
\end{equation}
The evolution of a state vector is described by the Schr\"{o}dinger type
equation
\begin{equation}
\idtau\psi=K\psi \label{eq:evolution}
\end{equation}
where for a free particle $K=\frac{P^{2}}{2M}$, resulting in the
\schstu equation (\ref{eq:sse}).

Horwitz and Arshansky \cite{HA82} have suggested a second order equation for
particles with \spinhalf. They argue as follows.
For a particle with spin the components of the
wave function must transform as a representation of the Lorentz group. The
norm
must be invariant, so the representation must be unitary. But the Lorentz
group
is a non-compact group, therefore the unitary representations are infinite
dimensional, containing all spins; such a ladder representation would,
however, introduce problems with
the application of the Pauli principle, for example, in the Sommerfeld model
of a metal.
If one were to use an induced representation based on the particle
four-momentum
as done by Wigner \cite{WIG49}, the expectation
value of $x^{\mu}$, which by Eq.\ (\ref{eq:canonvar}) is replaced by
$i\frac{\partial}{\partial p^{\mu}}$,
would not be covariant (the derivative acts on the unitary operator of the
little group defined by $p^{\mu}$).
Their solution consisted of introducing
a representation induced on the little group of a unit time-like vector, which
we denote here by $t^{\mu}$, which commutes
with $x^{\mu}$ and $p^{\mu}$. They described the transformation properties of
the wave function, and found the form of the positive definite, covariant norm
to be
\begin{equation}
N=\int d^{4}x\;\bar{\psi}_{\tau t}(x)\gamadt\psi_{\tau t}(x) \label{eq:norm}
\end{equation}
where $\psi_{\tau t}(x)$ is the Dirac spinor, and $\gamma^{\mu}$ the usual
Dirac matrices.
They constructed the Hermitian and anti-hermitian parts of the operator
$(\gamma\cdot P)$ under the scalar product associated with the norm. These
are, in Hermitian form (under the norm (\ref{eq:norm})),
\begin{eqnarray}
K_{L} & = & \frac{1}{2}((\gamma\cdot P)+(\gamma\cdot t)(\gamma\cdot P)
(\gamma\cdot t))=-(P\cdot t)(\gamma\cdot t) \\
K_{T} & = & \frac{1}{2}\gamaf((\gamma\cdot P)-(\gamma\cdot t)(\gamma\cdot P)
(\gamma\cdot t))=-2i\gamaf(P\cdot K)(\gamma\cdot t)  \label{eq:klkt}
\end{eqnarray}
where $K^{\mu}=\Sigma^{\mu\nu}t_{\nu}$ , $\Sigma^{\mu\nu}=
\frac{1}{4}i[\gamamu,\gamanu]$,
and the subscripts T and L denote transverse and longitudinal parts relative
to the time-like vector $t_{\mu}$. Since
\begin{equation}
\label{eq:k2}
\begin{array}{ccc}
K^{2}_{L}=(P\cdot t)^{2} & \;\; ; \;\; & K^{2}_{T}=P^{2}+(P\cdot t)^{2}
\end{array}
\end{equation}
for the equation of evolution, Eq.\ (\ref{eq:evolution}), one can write
\begin{equation}
\idtau\psi=\frac{1}{2M}(K^{2}_{T}-K^{2}_{L})\psi
\end{equation}
By introducing minimal coupling
$P_{\mu}\ \rightarrow P_{\mu}-eA_{\mu}$ they then obtained
\begin{equation}
\idtau\psi=\frac{(P-eA)^{2}}{2M}\psi+\frac{e}{2M}
\Sigma^{\mu\nu}_{t}F_{\mu\nu}\psi \label{eq:EMsecor}
\end{equation}
where
\begin{equation}
\Sigma^{\mu\nu}_{t}=\Sigma^{\mu\nu}+K^{\mu}t^{\nu}-K^{\nu}t^{\mu}
\end{equation}
This equation reproduces the correct gyromagnetic ratio, and does not contain
the non-hermitian
spin term appearing in Dirac's electromagnetically coupled, second order
equation
(in the special frame $t_{\mu}=(1,0,0,0)$, one easily sees that $\Sigma^{0j}$
is canceled so that there is no direct coupling of the electric field with
spin in this special frame). Note that $\Sigma^{\mu\nu}F_{\mu\nu}$,
appearing in the Dirac second order equation, contains $i{\bf \sigma\cdot E}$
as well as ${\bf \sigma\cdot H}$; the
former is not Hermitian in the norm $\int\psi^{\ast}(x)\psi(x)d^{3}x$, which
is Dirac's scalar product.

In a later work, Arensburg and Horwitz \cite{HAR90} extended the formalism to
a first order equation for \spinhalf. Since the $K_{T}$ part is responsible
for
the production of the correct gyromagnetic ratio, they postulated a first
order equation of the form
\begin{equation}
\idtau\psi=2(P\cdot K_{T})(\gamma\cdot t)\psi \label{eq:firstarens}
\end{equation}
(Their $K_{T}$ is equivalent to that of Ref. \cite{HA82} up
to a factor $-i\gamaf$). Furthermore, they found its solutions and associated
current,
and showed that the current, although exhibiting a space-like nature at each
point on the orbit (defined by $t_{\mu}$),
integrated over all possible $t_{\mu}$ in the forward light cone (completing
the natural scalar product of an induced representation),
reduces to a time-like current vector.

Eq.\ (\ref{eq:firstarens}) does not, however, conform to the \schstu equation
(\ref{eq:sse}) by iteration; it leads to a second order equation in $\tau$.
This is our original motivation in trying to obtain a new kind of first
order equation for \spinhalf particles in this framework.


\section*{III. General features of the equations}
\label{sec:genfeat}


\subsection*{3.1 Basic structure}
\label{subs:basicstruct}

In our attempt to find a manifestly covariant equation for an event with spin
we require some general features of the equation concerning the relation
between first and second order equations, namely the Dirac equation and the
Klein-Gordon equation. We proceed from the point of view that a first order
equation
is an additional condition on the second order one, while at the
same time introducing the notion of spin. In the process we
narrow down the number of available options and maintain only the suitable
ones.

The desired equation should contain first order derivatives only,
giving equal footing to the treatment of space and time.
Another requirement is that the equation be Lorentz
covariant, i.e., we wish the equation to be form covariant in respect to the
choice of inertial frames. We assume that the space-time derivatives
are coupled to an object constructed from $\gamma^{\mu}$ matrices and perhaps
some other four-vectors, and work with Dirac spinors, since we want the
theory to be as close as possible to the standard theory of Dirac.
For the actual
Lorentz transformations of the wave functions and equations we use the well
known
form \cite{BD,ITZUB} (this form is also applicable to the more general case
treated in \cite{HPR75}, \cite{HA82}; we use $\sigmunu$ in place of
$\Sigma^{\mu\nu}$ henceforth for notational simplicity)
\begin{equation}
S(\Lambda)=e^{-\frac{i}{4}\sigmunu\omegamunu } \label{eq:LorTran}
\end{equation}
where
\begin{equation}
\sigmunu=\frac{i}{2}[\gamma^{\mu},\gamma^{\nu}]
\end{equation}
and $\omegamunu$ are the antisymmetric transformation parameters.

We also require the equation to conform to a form of the \schstu equation by
iteration,
thus ensuring the free solutions of the spin particle to be also solutions of
that form of \schstu\/ equation, in parallel to the relation between Dirac's
equation and
the Klein-Gordon equation. This way each component of the wave function
satisfies
the free form of the \schstu\/ equation separately. From another point of view
we may regard the
desired equation as an additional condition on the solutions of the \schstu
equation, as the Dirac equation is to those of the Klein-Gordon equation.

Now we postulate the most general form of a first order spin equation
\begin{equation}
L(P)\psi=\sminus\nminus\idtau\psi +\splus M\nplus\psi \label{eq:genform}
\end{equation}
where $L(P)$ is a linear function of first order space-time derivatives,
namely a
function of $P_{\mu}$, and $\splus$ and $\sminus$ are sign variables to be
determined.
$\nplus$ and $\nminus$ are unknown matrices at this stage. The second term
on the
R.H.S. of Eq.\ (\ref{eq:genform}) has no derivatives. To compensate for the
dimensional
deficiency we introduce the scale factor M, which appears in the \schstu
equation. This term is necessary for maintaining the first
order derivative in respect to $\tau$, after the iteration is done.

To get the \schstu\/ equation we multiply by $L(P)$ on the left
\begin{equation}
L^{2}(P)\psi=\sminus L(P)\nminus\idtau\psi +\splus ML(P)\nplus\psi
\label{eq:ittgenform}
\end{equation}
Now, let us define the commutators
and anticommutators for the operators in Eq.\ (\ref{eq:genform})
\begin{eqnarray}
\antiCLp   & \equiv  &  \{L(P),\nplus\}           \nonumber \\
\antiCLn   & \equiv  &  \{L(P),\nminus\}          \nonumber \\
\antiCnn   & \equiv  &  \{\nplus,\nminus\}        \nonumber \\
\CLp       & \equiv  &  \mbox{[}L(P),\nplus]      \nonumber \\
\CLn       & \equiv  &  \mbox{[}L(P),\nminus]     \nonumber \\
\Cnn       & \equiv  &  \mbox{[}\nplus,\nminus]
\end{eqnarray}
We substitute Eq.\ (\ref{eq:genform}) into Eq.\ (\ref{eq:ittgenform}) to
obtain
\begin{eqnarray}
L^{2}(P)\psi & = & N_{-}^{2}\partial_{\tau}^{2}\psi -\spn M\antiCnn\idtau\psi
+\sminus\antiCLn\idtau\psi +\splus M\antiCLp\psi-M^{2}N_{+}^{2}\psi
\label{eq:itt2}
\end{eqnarray}

We do not want second order derivatives in respect to $\tau$, so we must have
\begin{equation}
N_{-}^{2}=0  \label{eq:npisnilp}
\end{equation}
Since we postulated that only first order space-time derivatives appear in
$L(P)$, and since Eq.\ (\ref{eq:itt2}) is supposed to coincide with the
\schstu\/ equation (\ref{eq:sse}), we require in addition to
Eq.\ (\ref{eq:npisnilp}) also that
\begin{equation}
\antiCLp =\antiCLn =0  \label{eq:antiCnilp}
\end{equation}
This gets rid of first order space-time derivatives in Eq.\ (\ref{eq:itt2}),
and leaves
only second order ones in $L^{2}(P)$. We still have an unwanted term,
$M^{2}N_{+}^{2}\psi$, and we require it to be zero (we could alternatively
absorb it in the phase of the wave function).
Thus, we obtain
\begin{equation}
[-\spn M\antiCnn]^{-1}L^{2}(P)\psi_{(\tilde{c})} =\idtau\psi_{(\tilde{c})}
\label{eq:ittshort}
\end{equation}
which is of the form of the desired equation (\ref{eq:sse}), if we can find
a solution for the conditions
\begin{eqnarray}
L^{2}(P)=\pm P^{2} \;\;  ; \;\;  \antiCnn =\pm 2
\end{eqnarray}
The subscript $(\tilde{c})$ in Eq.\ (\ref{eq:ittshort})
represents an equation derived using anticommutators.

We may use commutators instead of anticommutators, and repeating the previous
procedure
we reach the same conclusions concerning $\nplus$, $\nminus$ and $\antiCnn$.
However, this time there is the requirement that
\begin{equation}
\CLp =\CLn =0  \label{eq:Cnilp}
\end{equation}
The equation analogous to Eq.\ (\ref{eq:ittshort}) is
\begin{equation}
[\spn M\antiCnn]^{-1}L^{2}(P)\psi_{(c)} =\idtau\psi_{(c)}
\label{eq:ittshort2}
\end{equation}
The subscript $(c)$ represents an equation derived using commutators.
As we see later on, we shall need both forms.
In addition, we require that the equation have a continuity equation,
conserved positive definite probability, and currents.


\subsection*{3.2 Continuity equation and currents}
\label{subs:contcurr}

We are interested in achieving a continuity equation for
Eq.\ (\ref{eq:genform}) of the form
\begin{eqnarray}
\partial_{\alpha}j^{\alpha}=0 \;\; ;  \;\; (\alpha =0,1,2,3,\tau)
\label{eq:gencont}
\end{eqnarray}
in order to obtain the usual interpretation, where $j^{\tau}=\rho$ should
be the
probability density, conserved through Eq.\ (\ref{eq:gencont}).
The simple way to do this is clearly shown in all textbooks
\cite{BD,ITZUB,Schweb} for the Dirac equation, and we follow this general
method.
We observe that since $\nminus$ is nilpotent, the probability density
cannot be
of the usual Dirac form $\psidag\psi$, but contains a matrix between
$\psidag$ and $\psi$ (a nilpotent does not have an inverse so there is no
way to get a pure $\idtau\psi$ term in Eq.\ (\ref{eq:genform})).
Refs. \cite{HA82,HAR90} show that the scalar product for the second order
equation
obtained from Eq.\ (\ref{eq:norm}) is
\begin{equation}
(\phi_{Dirac},\psi_{Dirac})=\int d^{4}p\;\bar{\phi}_{Dirac}\;(\gamma\cdot t)\;
\psi_{Dirac} \label{eq:oursp}
\end{equation}
where $t_{\mu}$ is the same time-like four-vector with norm $t^{2}=\mbox{-}1$
mentioned
earlier. The time-like nature of $t_{\mu}$ is crucial for the scalar product
to be
positive definite. [Of course, $\phi$, $\psi$ depend on $t^{\mu}$,
so that (\ref{eq:oursp}) is a covariant structure on a bundle (i.e., as an
induced representation).]
The probability density
is just the special case of the integrand of the scalar product when
$\phi =\psi$.
To find the continuity equation one multiplies the equation from the left
by $\psidag\gamaz$ to get
\begin{equation}
\psibar L(P)\psi =\sminus\psibar\nminus\idtau\psi+\splus M\psibar\nplus\psi
\label{eq:psifleft}
\end{equation}
then multiply the conjugated equation from the right by $\gamaz\psi$ to get
\begin{equation}
\psidag L^{\dagger}(-\stackrel{\leftarrow}{P})\gamaz\psi =-\sminus\psidag
i\stackrel{\leftarrow}{\dtau}N_{-}^{\dagger}\gamaz\psi +
\splus M\psidag N_{+}^{\dagger}\gamaz\psi \label{eq:psifright}
\end{equation}
and subtract Eq.\ (\ref{eq:psifright}) from Eq.\ (\ref{eq:psifleft}).
In order to achieve
the form (\ref{eq:gencont}) we must have
\begin{eqnarray}
\gamaz(\nplus)^{\dagger}\gamaz & = & \nplus    \nonumber \\
\gamaz(\nminus)^{\dagger}\gamaz & = & \nminus  \\
\gamaz(L^{\dagger}(-\stackrel{\leftarrow}{P}))\gamaz & = &
-L(\stackrel{\leftarrow}{P}) \nonumber
\end{eqnarray}
so that we get
\begin{equation}
\psibar(L(\stackrel{\rightarrow}{P})+L(\stackrel{\leftarrow}{P}))\psi =
\sminus\psibar(\nminus[i\stackrel{\leftarrow}{\dtau}+i\stackrel{\rightarrow}
{\dtau}])\psi
\end{equation}
and finally
\begin{equation}
\dmu(\psibar b^{\mu}\psi)=\sminus\dtau(\psibar\nminus\psi) \label{eq:speccont}
\end{equation}
where it is assumed that the form of $L(P)$ is $b^{\mu}i\dmu$, and
$b^{\mu}$ is some yet unspecified object with a Lorentz index which
couples to $P_{\mu}$. We must therefore now search for nilpotents for which
$(\psibar\nminus\psi)$ is positive. A general analysis is carried in Appendix
\ref{app:all_potents}.

\subsection*{3.3 Acceptable nilpotents}
\label{subs:acceptnils}

One finds (in Appendix \ref{app:all_potents}) that the candidates for
nilpotents
which pertain to a positive definite probability density are:
\begin{equation}
\label{eq:availnils}
\begin{array}{lcl}
(\gamma\cdot l^{(\pm)})    & : & \gamadt({\bf 1}\pm\gamaf) \\ \\
(\gamma\cdot l^{(\pm)})    & : & \gamadt\pm\sigmats \\ \\
\gamadt\pm i\gamaf & : & \gamadt\pm\sigmats
\end{array}
\end{equation}
where we denote
\begin{equation}
\sigmats\equiv \sigma^{\mu\nu}t_{\mu}s_{\nu}
\end{equation}
and $l$, $s$, and $t$ are light-like, space-like, and time-like vectors
respectively.

Nilpotents of this type come in {\sl non-equivalent} pairs, in the sense
that no
unitary transformation connects the members of the pair (see Appendix
\ref{app:all_potents}).
Furthermore, pairs in the {\sl columns} of Eqs.\ (\ref{eq:availnils})
may be {\sl equivalent}
since we are able to transform from one to another through a unitary
transformation.
In any case we look at all four possibilities in the quest for $L(P)$. Each
nilpotent pair can be seen to be formed of two parts, a projection operator
$Pr_{\pm}$, and
$\gdt$, such that $\gdt\cdot Pr_{-}=Pr_{+}\cdot\gdt$.
In particular,
\begin{equation}
\label{eq:availnils2}
\begin{array}{lcl}
(\gamma\cdot l^{(\pm)}) & = & \gdt({\bf 1}\pm\gdt(\gamma\cdot s)) \\ \\
\gdt\pm i\gamaf       & = & \gdt({\bf 1}\pm i\gdt\gamaf) \\ \\
\gdt\pm\gdt\gamaf     & = & \gdt({\bf 1}\pm\gamaf) \\ \\
\gdt\pm\sigmats       & = & \gdt({\bf 1}\pm i(\gamma\cdot s))
\end{array}
\end{equation}
The last of Eqs.\ (\ref{eq:availnils2}) is due to the fact that $(t\cdot s)=0$
and $\sigmats=\frac{i}{2}[\gdt,\gds]$.


\subsection*{3.4 The four-momentum part : $L(P)$}
\label{subs:fourmom}

We wish $L(P)$ to be linear in first derivatives, have zero anticommutators
with $N_{\pm}$ (see Eq.\ (\ref{eq:antiCnilp})), produce the four-current part
of the continuity equation, iterate to $\pm P^{2}$, and be Lorentz covariant.
We check the four available forms of nilpotents in Eqs.\ (\ref{eq:availnils}),
stemming
from the specific reference frame in which they were found. To comply with
Eq.\ (\ref{eq:antiCnilp}) we search for all $\Gamma$'s (see Appendix
\ref{app:all_potents})
which anticommute with
$N_{\pm}$, then we couple $P_{\mu}$ in all possible ways, and
check the anticommutators, and other criteria. The same procedure must be
repeated
for commutators of Eq.\ (\ref{eq:Cnilp}). Going through all options of
Eq.\ (\ref{eq:availnils2}) is a
tedious process, and we shall demonstrate only an example. Let us partly
analyze the nilpotent form of $N_{\pm}\equiv(\gamma\cdot l^{(\pm)})$.
As the original non-covariant form we take $\gamaz\pm\gamaone$. The
$\Gamma$'s that anticommute
with $\gamaz =\Gamma_{2}$ and $\gamaone =-i\Gamma_{3}$ are :
\begin{equation}
\Gamma_{4},\Gamma_{5},\Gamma_{6},\Gamma_{16} \;\; \Longleftrightarrow \;\;
i\gamatwo ,i\gamatre ,\gamaz\gamaone ,\gamaf \label{eq:firstnil}
\end{equation}
Since it is required that $L^{2}(P)$ must be a multiple of the identity
matrix,
if we wish to put in any
combination of the matrices appearing in Eq.\ (\ref{eq:firstnil}), we must
divide them
into two sets, for which the matrices in a set anticommute among themselves,
and commute with all other matrices of the other set, and consider each set
separately. Otherwise,
after the iteration of $L(P)$ we would find non-diagonal parts due to the
anticommutator of two commuting matrices. The two sets
are $(\gamaz\gamaone)$ and $(i\gamatwo ,i\gamatre ,\gamaf)$.
For example, the combination of $\gamaz\gamaone$ can be coupled to
$P^{\mu}$ in a covariant way as
\begin{equation}
\gamaz\gamaone P_{0}v_{1}\pm \gamaz\gamaone P_{1}v_{0} \rightarrow \sigmapv
\end{equation}
where $P_{\mu}=(P_{0},P_{1},0,0)$, and $v=(v_{0},v_{1},0,0)$ is some vector,
in this frame (since there are two matrices we form a tensor). If we want to
conform to Eq.\ (\ref{eq:antiCnilp}) by taking $(\gamma\cdot l^{(\pm)})$ as
$\nplus,\nminus$
and we do not want $P_{\mu}$ to be only light-like, then $v$ must be
light-like
(since $\tilde{C}_{l\pm}=\{\sigma Pv,\gamma\cdot l^{(\pm)}\}=0$ implies
$P\propto l^{(\pm)}$
or $v\propto l^{(\pm)}$). But then the vector $l^{(\pm)}$ must be the same in
both $\nplus,\nminus$
(either $l^{(+)}$ or $l^{(-)}$), which makes $\antiCnn=0$. This contradicts
our basic assumptions, so this sub-form must be discarded.

We have required in Section III that the iteration of the first
order equation
should give us $L^{2}(P)=\pm P^{2}$.
Checking all available options of the nilpotents of
Eq.\ (\ref{eq:availnils2}),
indicates that we are unable to get such a form by iteration
(at least not under our initial assumptions
concerning the iteration process). [In Section VIII we shall
show how to produce $P^{2}$ by
a somewhat different iteration procedure (obtained in a rigid limiting case);
in that procedure we lose
the Hermiticity of the spin term in the second order electromagnetically
coupled
equation, getting an expression as in the Dirac equation. Furthermore, we lose
the structure of the continuity equation.] This applies to
commutators and anticommutators alike. Nevertheless we proceed, considering
the transverse and longitudinal options separately, combining them in Section
 VI. We start by choosing from the anticommutator forms.
From them we choose
\begin{equation}
N_{\pm}=\gdt({\bf 1}\pm\gamaf)
\end{equation}
and
\begin{equation}
L_{\perp}(P)=\sigma^{\mu\nu}P_{\mu}t_{\nu}
\end{equation}
This is because in doing so, we deal with
only one additional four-vector, $t_{\mu}$, while retaining the ability to
achieve the gyromagnetic ratio, and exhibiting some features concerning
chirality due to the use of the projection operators
$\frac{1}{2}({\bf 1}\pm\gamaf)$.
It also seems that this is the simplest choice which has as many benefits as
possible
in this situation. As we show later on, this choice gives rise to the
transverse
\schstu equation. Not excluding the existence of a longitudinal equation
also, we
follow our current choice for $N_{\pm}$, and take the simplest choice for the
longitudinal equation, namely
\begin{equation}
L_{\parallel}(P)=i(P\cdot t)
\end{equation}


\section*{IV. The equations of motion}
\label{sec:eqofmotion}

We now discuss the consequences of the above choice for the forms of
$N_{\pm}$ and $L(P)$'s.
This discussion concerns what we call the transverse and longitudinal \schstu
equations.


\subsection*{4.1 Transverse equations of motion}
\label{subs:Teqofmotion}

We have
\begin{equation}
\antiCnn=\{\gdt\pminus,\gdt\pplus\}=-4t^{2}
\end{equation}
taking $t^{2}=-1$, a unit time-like four-vector, we have to arrange an
additional
factor of $2$ to get the desired form of $\frac{1}{2M}$. We place this factor
with $\nplus$. [In Section VIII we show
how to get the solutions of the Dirac equation in the limit where $p_{\mu}$
and $t_{\mu}$ coincide.
For that to happen we have to take the factor $2$ as indicated.]
The anticommutators satisfy of course
\begin{equation}
\{\gdt({\bf 1}\pm\gamaf),\sigmapt\}=0
\end{equation}
which follows from the important relation
\begin{equation}
\{\gdt,\sigmapt\}=0 \label{eq:anticom1}
\end{equation}
Iterating $L_{\perp}(P)$ we obtain
\begin{equation}
\sigmapt^{2}=P^{2}t^{2}-(P\cdot t)^{2} \label{eq:sqrsigmapt}
\end{equation}
We define the transverse and longitudinal momentum, relative to $t_{\mu}$, as
\begin{eqnarray}
P_{\perp\mu}     & = & P_{\mu}+(P\cdot t)t_{\mu} \label{eq:Tmomet} \\
P_{\parallel\mu} & = & -(P\cdot t)t_{\mu} \label{eq:Lmomet}
\end{eqnarray}
and we therefore have
\begin{displaymath}
\sigmapt^{2}=t^{2}P^{2}_{\perp}=-P^{2}_{\perp}
\end{displaymath}
The equation of motion is
\begin{equation}
-\hbar c(\sigma^{\mu\nu}i\partial_{\mu}t_{\nu})\psi_{\perp}=\hbar\sminus\gdt
\pminus\idtau\psi_{\perp}+\splus\frac{Mc^{2}}{2}\gdt\pplus\psi_{\perp}
\label{eq:eqtn1}
\end{equation}
We display $\hbar$ and $c$ in this
principal equation; elsewhere we take $\hbar=c=1$. It is understood that
$x^{0}\equiv ct$.
Notice that in our metric $P_{\mu}=-i\frac{\partial}{\partial x^{\mu}}$.
\par Iterating, we obtain
\begin{displaymath}
\frac{P^{2}_{\perp}}{\splus\sminus 2M}\psi_{\perp}=\idtau\psi_{\perp}
\end{displaymath}
This forces the signs (with our choice of phase) to be either both positive
or both negative, so that \mbox{$\splus\sminus=1$}, and we get the
{\sl transverse}
\schstu equation
\begin{equation}
\frac{P^{2}_{\perp}}{2M}\psi_{\perp}=\idtau\psi_{\perp}  \label{eq:transsse}
\end{equation}
which we assume to hold. Using the notation of Horwitz and Arshansky
\cite{HA82}, \\
$L^{2}(P)=-K^{2}_{T}$, which generates the evolution of transverse momentum
only.
Note that $\psi_{\perp}$ contains only the transverse space-time coordinates.

We have not excluded the possibility of a longitudinal
\schstu equation; from another, independent, first order equation, which we
shall discuss later, we shall also have

\begin{equation}
\frac{P^{2}_{\parallel}}{2M}\psi_{\parallel}=\idtau\psi_{\parallel}
\label{eq:longsse}
\end{equation}
where $\psi_{\parallel}$ contains only longitudinal space-time coordinates.
The subscripts $\perp$ and $\parallel$ are introduced for making a distinction
between the
wave functions of the transverse and longitudinal equations. These indeed
constitute kinematically independent degrees of freedom.
\par Since Eq.\ (\ref{eq:eqtn1}) contains projectors we can decompose it
 into two
 coupled equations by multiplying it from the left with the same projectors.
Denoting the wave function as composed of two chiral spinors, (two-component
spinors in the chiral representation),
\begin{eqnarray}
\psi_{L}=\frac{1}{2}\pminus\psi & & \psi_{R}=\frac{1}{2}\pplus\psi
\label{eq:twoprojs}
\end{eqnarray}
we obtain
\begin{eqnarray}
\gdt\sigmapt\psiLp & = & \splus M\psiRp \label{eq:lr1} \\
\gdt\sigmapt\psiRp & = & 2\sminus\idtau\psiLp \label{eq:rl1}
\end{eqnarray}
This is an explicit chiral decomposition, not symmetric for left and right
handed
spinors, a matter discussed later on. Such a form for equations of the
first order in $\tau$-formalism has been proposed by Davidon \cite{DAV55}, in
a somewhat {\sl ad hoc} manner. He could not overcome the problem of
finding a positive definite probability density, and he obtained a
non-hermitian
spin term after coupling the electromagnetic field. The equations
(\ref{eq:lr1}),(\ref{eq:rl1})
are completely equivalent to the one equation (\ref{eq:eqtn1}); it is
just a rewriting.

Concerning the sign variables $\splus$ and $\sminus$, it seems that we have
two choices. Denoting by $\psi_{\perp +}$ and $\psi_{\perp -}$ the solutions
of the two sign
versions of Eq.\ (\ref{eq:eqtn1}), by adding and subtracting the
two versions, we get exactly the equations (\ref{eq:lr1}),(\ref{eq:rl1}), with
the following relation between $\psi_{\perp +}$,$\psi_{\perp -}$ and
$\psi_{\perp L}$,$\psi_{\perp R}$
\begin{eqnarray}
\psi_{\perp +}+\psi_{\perp -}=\psiRp \\
\psi_{\perp +}-\psi_{\perp -}=\psiLp
\end{eqnarray}
As can be seen from Eq.\ (\ref{eq:speccont}),
the sign $\sminus$ (equal to $\splus$ in our treatment), sets the sign of
the so called $g_{\tau\tau}$ component of the metric.

After we have discussed the two options of sign assignment, we cannot ignore
the two possibilities of nilpotent assignment. Instead of deducing
Eq.\ (\ref{eq:eqtn1}),
we could equally well introduce a complementary equation with the places
of the nilpotents interchanged
\begin{equation}
-\sigma^{\mu\nu}i\partial_{\mu}t_{\nu}\varphi_{\perp}=
\sminus\gdt\pplus\idtau\varphip+\splus\frac{M}{2}\gdt\pminus\varphi_{\perp}
\label{eq:eqtn2}
\end{equation}
where we left the signs in place. Eq.\ (\ref{eq:eqtn2}) has all the
features of
Eq.\ (\ref{eq:eqtn1}), except that the role of the left and right handed
spinors
is interchanged. Writing the two relevant coupled equations analogous to
Eqs.\ (\ref{eq:lr1}),(\ref{eq:rl1})
we get
\begin{eqnarray}
\gdt\sigmapt\varphiRp & = & \splus M\varphiLp \label{eq:lr2} \\
\gdt\sigmapt\varphiLp & = & 2\sminus\idtau\varphiRp \label{eq:rl2}
\end{eqnarray}

Now, if we abide by Eq.\ (\ref{eq:eqtn1}), and describe the theory as
portraying the evolution of a left-handed spinor, $\psi_{\perp L}$, while
$\psi_{\perp R}$ is just a sort
of auxiliary field, we give up right-handed events. This cannot be
done without some justification. In Section \ref{sec:probcurr} we see that
such an equation
gives rise to a continuity equation, (through a simple procedure similar to
the one performed for the Dirac equation), where the probability density
is composed
of left-handed spinors only.

The interpretation of $\psi_{\perp R}$ as an auxiliary field,
justified by Eqs.\ (\ref{eq:rl1}),(\ref{eq:lr1})
where only $\psi_{\perp L}$ is seen to evolve according to $\tau$, the
existence of
the other equally justified equation (\ref{eq:eqtn2}), the inability to prefer
Eq.\ (\ref{eq:eqtn1}) over Eq.\ (\ref{eq:eqtn2}), and the requirement for
a reasonable probability density, lead us to a unification of the two.

We define
\begin{eqnarray}
\phip\equiv\psiLp+\varphiRp & & \chip\equiv\psiRp+\varphiLp
\end{eqnarray}
These are now four-spinors, where $\chip$ is an auxiliary field, and $\phip$
is the main field, evolving with $\tau$. The two sets of coupled equations
(\ref{eq:lr1}),(\ref{eq:rl1}) and (\ref{eq:lr2}),(\ref{eq:rl2}), become
\begin{eqnarray}
\gdt\sigmapt\phip & = & \splus M\chip \label{eq:phichi} \\
\gdt\sigmapt\chip & = & 2\sminus\idtau\phip \label{eq:chiphi}
\end{eqnarray}
by just adding them up. We denote them as the ``extended'' form.
We therefore have achieved a satisfactory probability density
of four-spinors, which is sufficient for the two irreducible representations
of $SL(2,C)$ (see \cite{HA82},\cite{HAR90}).
Of course by iteration we get the transverse \schstu equation for $\phip$ and
for $\chip$.


\subsection*{4.2 Longitudinal equations of motion}
\label{subs:Leqofmotion}

In dealing with a first order equation that agrees with the longitudinal
\schstu
equation, Eq.\ (\ref{eq:longsse}), we check it for the nilpotents and momentum
part we choose. Since the nilpotents are the same as for the transverse
version all conclusions from Subsection \ref{subs:Teqofmotion} apply here too.
Iterating $L_{\parallel}(P)$ we get $P^{2}_{\parallel}=-(P\cdot t)^{2}$, and
the relation
between $\splus$ to $\sminus$ is maintained as in the transverse case. The
equation of motion is
\begin{equation}
i(-i\partial_{\mu}t^{\mu})\psi_{\parallel}=\sminus\gdt\pminus\idtau
\psi_{\parallel}+\splus\frac{M}{2}\gdt\pplus\psi_{\parallel} \label{eq:leqtn1}
\end{equation}
Using projectors of Eq.\ (\ref{eq:twoprojs}) we can decompose it in two
\begin{eqnarray}
i\gdt(P\cdot t)\psiLl & = & \splus M\psiRl \label{eq:llr1} \\
i\gdt(P\cdot t)\psiRl & = & 2\sminus\idtau\psiLl \label{eq:lrl1}
\end{eqnarray}
The discussion concerning both options in using the sign variables
applies here as well, and we do not repeat it. As for the transverse case,
Eqs.\ (\ref{eq:llr1}),(\ref{eq:lrl1}) can be used to show that the left and
right handed spinors pertain to the longitudinal \schstu equation.

Exchanging the place of nilpotents in the equation to get the complementary
equations of motion for the right handed part as the main field, while
maintaining
the interpretation of the auxiliary field, and adding the two types of
equations as in the transverse case, leads to the coupled equations
\begin{eqnarray}
i\gdt(P\cdot t)\phil & = & \splus M\chil \label{eq:lphichi} \\
i\gdt(P\cdot t)\chil & = & 2\sminus\idtau\phil \label{eq:lchiphi}
\end{eqnarray}
with the same structure of $\phil$ and $\chil$ as in the transverse case.

\section*{V. Probability density and currents}
\label{sec:probcurr}

The continuity equations for the transverse and longitudinal versions are
obtained by using the procedure for obtaining the primary form of a
continuity equation in Subsection \ref{subs:contcurr}. We show that the
currents form consistent continuity equations, and are related to the
Gordon decomposition of the Dirac current.

\subsection*{5.1 Currents of the transverse equation}
\label{subs:Tprobcurr}

Using the procedure in Subsection \ref{subs:contcurr} on
Eq.\ (\ref{eq:eqtn1}),
we obtain
\begin{equation}
-\dmu(\psibarp\sigmunu t_{\nu}\psip)=2\sminus\dtau(\psibarp\gdt
\frac{1}{2}\pminus\psip)  \label{eq:primcont}
\end{equation}
We observe that the R.H.S. of Eq.\ (\ref{eq:primcont}) is a probability
density, {\sl exactly}
in the sense defined by Horwitz and Arshansky \cite{HA82}, but achieved
{\sl without}
explicit group theoretical arguments. It is obviously positive definite
for wave functions containing left handed components, which
can be seen by transforming to a reference frame where $t_{\mu}=(1,0,0,0)$.
We denote this probability
density (in the chiral representation of $\gamma$ matrices) as
\begin{equation}
\rho_{\perp L}\equiv(\psibarp\gdt\frac{1}{2}\pminus\psip)=\psidag_{\perp L}
(\tilde{\sigma}^{\mu}t_{\mu})\psiLp
\end{equation}
where
\begin{eqnarray}
\tilde{\sigma}^{\mu}\equiv({\bf 1},-\sigma^{i}) & \;\; ; \;\; & \sigma^{\mu}
\equiv({\bf 1},\sigma^{i})
\end{eqnarray}
and $\sigma^{i}$ are the Pauli matrices. The appearance of $\rho_{\perp L}$
only in
Eq.\ (\ref{eq:primcont}) is the chief reason for introducing both
Eqs.\ (\ref{eq:eqtn1}),(\ref{eq:eqtn2}).
From Eq.\ (\ref{eq:eqtn2}) we obtain in the same way $\rho_{\perp R}$, and
we can add them
together, and obtain $\rho_{\perp}=\rho_{\perp L}+\rho_{\perp R}$, thus
having the a density containing both $SL(2,C)$
representations as in \cite{HA82,HAR90}. Since $\rho_{\perp R}$ is also
positive definite, so is $\rho_{\perp}$, as required.

We define the four-currents in Eq.\ (\ref{eq:primcont}) as
\begin{equation}
j^{\mu}_{\perp L}\equiv\psibarp(\sigmunu t_{\nu})\psip \label{eq:jmuL}
\end{equation}
so we actually have
\begin{equation}
-\dmu j^{\mu}_{\perp L}=2\sminus\dtau\rho_{\perp L}
\end{equation}
which is the five dimensional conservation theorem required. Let us consider
\begin{eqnarray}
\bar{\psi}_{\perp L}\sigmapt\psiRp & = & 2\sminus\bar{\psi}_{\perp L}\gdt
\idtau\psiLp \label{eq:lbarl}
\end{eqnarray}
and the conjugate of Eq.\ (\ref{eq:lbarl})
\begin{eqnarray}
\bar{\psi}_{\perp R}\sigmaptop\psiLp & = & 2\sminus(\idtau
\bar{\psi}_{\perp L})\gdt\psiLp \label{eq:lbarlhc}
\end{eqnarray}
Using the equations of motion, Eqs.\ (\ref{eq:lr1}),(\ref{eq:rl1}), we find
\begin{equation}
\label{eqs:LtoR}
\begin{array}{lcl}
\dtau\psiRp    & = & \frac{1}{\splus M}\gdt\sigmapt(\dtau\psiLp) \\ \\
\dtau\psiRbarp & = & \frac{1}{\splus M}(\dtau\psiLbarp)\gdt\sigmaptop \\ \\
\dtau\psiLp    & = & -\frac{i}{2\sminus}\gdt\sigmapt\psiRp \\ \\
\dtau\psiLbarp & = & \frac{i}{2\sminus}\psiRbarp\gdt\sigmaptop
\end{array}
\end{equation}
Using Eq.\ (\ref{eq:lr1}), and the sum of Eqs.\ (\ref{eq:lbarl}),
(\ref{eq:lbarlhc}) we get
\begin{eqnarray}
\partial_{\tau}\rho_{\perp L} & = & -\frac{i}{2M}\psiLbarp(\sigmapt\gdt
\sigmapt+\gdt\sigmaptop\sigmaptop)\psiLp \nonumber \\
 & = & \frac{i}{2M}\psiLbarp\gdt(-\stackrel{\rightarrow}{P_{\perp}^{2}}
+\stackrel{\leftarrow}{P_{\perp}^{2}})\psiLp
\end{eqnarray}

To simplify the expression for $\rho_{\perp L}$ we state some useful relations
\begin{eqnarray}
P_{\perp\mu}   &  \equiv  &   -i\partial_{\perp\mu}\equiv -i\dmu+
(-i\dnu\cdot t^{\nu})t_{\mu}  \nonumber \\
P^{2}_{\perp}  &    =     &   -\partial_{\perp\mu}\partial^{\mu}_{\perp}
\end{eqnarray}
In the reference frame where
$t_{\mu}=(1,0,0,0)$ the transverse derivative is seen to be
\begin{eqnarray}
\partial_{\perp 0}=0 \nonumber \\
\partial_{\perp i}=\partial_{i}
\end{eqnarray}
So,
\begin{equation}
\partial_{\tau}\rho_{\perp L}=\frac{i}{2M}\partial_{\perp\mu}(\psiLbarp\gdt
\stackrel{\leftrightarrow}{\partial_{\perp}^{\mu}}\psiLp) \label{eq:conttranL}
\end{equation}
Using Eqs.\ (\ref{eqs:LtoR}) we find that the $\tau$ derivative of
the auxiliary
field also obeys a continuity equation, exactly as the main field with
$\psiLp$ exchanged with $\psiRp$
\begin{equation}
\partial_{\tau}\rho_{\perp L_{aux}}=\frac{i}{2M}\partial_{\perp\mu}(\psiRbarp
\gdt\stackrel{\leftrightarrow}{\partial_{\perp}^{\mu}}\psiRp)
\label{eq:conttranLa} \label{eq:rholperp}
\end{equation}
where we therefore understand $\rho_{\perp L_{aux}}$ as a
``density'' associated with the auxiliary field of $\psiLp$, namely with
$\psiRp$.
This form of current resembles the Klein-Gordon current, and when taken
in a frame
where $t_{\mu}=(1,0,0,0)$ we recover the space part of the current.

To make a connection with the form of currents exhibited by the Dirac
equation, we re-express Eq.\ (\ref{eq:primcont}) as
\begin{equation}
-t_{\nu}\tilde{j}^{\nu}_{\perp L}=2\sminus t_{\mu}\dtau
\tilde{\rho}^{\mu}_{\perp L} \label{eq:decform}
\end{equation}
where
\begin{eqnarray}
\tilde{j}^{\nu}_{\perp L} & \equiv & \dmu(\psibarp\sigmunu\psip) \\
\tilde{\rho}^{\mu}_{\perp L} & \equiv & \psiLbarp\gamamu\psiLp
\end{eqnarray}
Now we integrate over a space-like hypersurface enclosing a four-dimensional
volume,
where $t_{\mu}$ is normal to the surface (see, for example, \cite{Schw51}).
Therefore, we imagine a space-like hypersurface perpendicular to $t_{\mu}$
and another one like it some distance above, meeting it only at infinity;
it then follows from Eq.\ (\ref{eq:decform}) that
\begin{equation}
-\int_{\Sigma_{t}}d\Sigma\; t_{\nu}\tilde{j}^{\nu}_{\perp L}=2\sminus\dtau
\int_{\Sigma_{t}}d\Sigma\; t_{\mu}\tilde{\rho}^{\mu}_{\perp L}
\end{equation}
where $\Sigma_{t}$ denotes the closed space-like hypersurface. We transform
to the four-divergence
\begin{equation}
-\int_{V_{t}}dv \;\dnu\tilde{j}^{\nu}_{\perp L}=2\sminus\dtau\int_{V_{t}}dv
\;\dmu\tilde{\rho}^{\mu}_{\perp L} \label{eq:dirspincur}
\end{equation}
But since
\begin{equation}
-\dnu\tilde{j}^{\nu}_{\perp L}=\dnu\dmu(\psibarp\sigmunu\psip)=0
\label{eq:GD1}
\end{equation}
on account of the antisymmetry of $\sigmunu$ and the symmetry of $\dmu\dnu$,
we get
\begin{equation}
\int_{V_{t}}dv \;\dtau\dmu\tilde{\rho}^{\mu}_{\perp L}=0
\end{equation}
Now, assuming this is true for the integrand also, i.e., for any thin slice
$V_{t}$, we obtain
\begin{equation}
\dtau\dmu(\psiLbarp\gamamu\psiLp)=0
\end{equation}
and hence
\begin{equation}
\dmu(\psiLbarp\gamamu\psiLp)=f(x^{\mu}) \label{eq:pseudoDL}
\end{equation}
The quantity $\dmu(\psiLbarp\gamamu\psiLp)$ is independent of $\tau$, and if
it is not zero, it is some $\tau$-independent function $f(x^{\mu})$.

Comparing to the Gordon decomposition of the Dirac currents one finds a
similar structure when applying the continuity equation.

We now consider the combination of the above methods and equations
concerning Eq.\ (\ref{eq:eqtn1}), with their application to the nilpotent
interchanged equation, Eq.\ (\ref{eq:eqtn2}).
The same line of thought, when applied to Eq.\ (\ref{eq:eqtn2}), yields
the same
results, only $\varphiRp$ replaces $\psiLp$, $\varphiLp$ replaces $\psiRp$,
$\dtau\rho_{\perp R}$ replaces $\dtau\rho_{\perp L}$.
Ultimately we get
the continuity equation for the main field
\begin{equation}
\dtau(\phibarp\gdt\phip)=\frac{i}{2M}\partial_{\perp\mu}
(\phibarp\gdt\stackrel{\leftrightarrow}{\partial_{\perp}^{\mu}}\phip)
\label{eq:fincurrs}
\end{equation}
Furthermore, one obtains for the total main field, a $\tau$ independent
continuity equation
resembling the one in Dirac's theory
\begin{equation}
\dtau\dmu(\phibarp\gamamu\phip)=0
\end{equation}
If $\dmu(\phibarp\gamamu\phip)=0$, (i.e., in case the sum of divergences for
$\psi_{\perp L}$ and $\varphi_{\perp R}$ cancel),
we have the form of the Dirac continuity equation, but for the transverse
field (recall that the wave function $\phip$ itself still has dependence
on $t_{\mu}$).

We have
a physical behavior of the particles described by our equations, which
resembles
the physical behavior of Dirac's particles, without the use of Dirac's
equation, but using instead the transverse equation. It is interesting to
observe that the Dirac type currents were
obtained from the {\sl probability density} of our theory. We may view this as
though our theory is in some sense a fundamental underlying structure for
what is essentially Dirac's theory.

We remark, moreover, that we have obtained in our method, a scalar product
which agrees with that of \cite{HA82}, i.e.,
\begin{equation}
(\phi_{\perp 1},\phi_{\perp 2})=\int d^{4}x\;\bar{\phi}_{\perp 1}\;
\gdt\phi_{\perp 2} \label{eq:myscalar}
\end{equation}

\subsection*{5.2 Currents of the longitudinal equation}
\label{subs:Lprobcurr}

Using the procedure in Subsection \ref{subs:contcurr} on
Eq.\ (\ref{eq:leqtn1}),
we obtain
\begin{equation}
\psibarl\stackrel{\leftrightarrow}{\dmu}t^{\nu}\psil=2\sminus
i\dtau(\psibarl\gdt\frac{1}{2}\pminus\psil)  \label{eq:lprimcont}
\end{equation}
In the R.H.S. we identify a positive definite quantity. Unfortunately, we
cannot
understand Eq.\ (\ref{eq:lprimcont}) as a continuity equation. However, we can
overcome the problem by observing similar features of the equations as in the
transverse case. We state some features of the longitudinal momentum
\begin{eqnarray}
P_{\parallel\mu} & \equiv & -i\partial_{\parallel\mu}\equiv
-(-i\partial_{\nu}t^{\nu})t_{\mu} \nonumber \\
P^{2}_{\parallel}  &    =     &   -\partial_{\parallel\mu}
\partial^{\mu}_{\parallel} \nonumber \\
\end{eqnarray}
and in the special frame where $t_{\mu}=(1,0,0,0)$ we have
\begin{eqnarray}
\partial_{\parallel 0} & = & \partial_{0} \nonumber \\
\partial_{\parallel i} & = & 0
\end{eqnarray}
For a treatment analogous to that of the transverse equation we find
\begin{equation}
\partial_{\tau}\rho_{\parallel L}=\frac{i}{2M}\partial_{\parallel\mu}
(\psiLbarl\gdt\stackrel{\leftrightarrow}{\partial_{\parallel}^{\mu}}\psiLl)
\label{eq:lconttranL}
\end{equation}
Using the space-like hypersurface integration while defining
\begin{eqnarray}
\tilde{j}^{\nu}_{\parallel L}         & \equiv &
\psibarl\stackrel{\leftrightarrow}{\partial^{\mu}}\psil \nonumber \\
\tilde{\rho}^{\mu}_{\parallel L}      & \equiv &
\psiLbarl\gamamu\psiLl \nonumber \\
t_{\nu}\tilde{j}^{\nu}_{\parallel L} & = &
2\sminus t_{\mu}i\dtau\tilde{\rho}^{\mu}_{\parallel L}
\end{eqnarray}
we get
\begin{equation}
\int_{V_{t}}dv \;2\sminus\dtau\dmu(\psiLbarl\gamamu\psiLl)=-\int_{V_{t}}dv
\;\dnu(\psibarl i\!\stackrel{\leftrightarrow}{\partial^{\nu}}\!\!\psil)
\label{eq:ld1}
\end{equation}

Eq.\ (\ref{eq:ld1}) is similar to Eq.\ (\ref{eq:dirspincur}) in the sense that
the probability density gives rise to a form of a Dirac current.
The transverse current of Eq.\ (\ref{eq:dirspincur}) has the form of the spin
current
of the Gordon decomposition of the Dirac current, while the longitudinal
current
of Eq.\ (\ref{eq:ld1}) has the form of the convection part. It seems that
the two parts of the Gordon decomposition manifest themselves in the two
versions of the equations of motion.
In case
$\dmu(\psiLbarl\gamamu\psiLl)$ is independent of $\tau$, one would have
a result for \\
$\dnu(\psibarl i\!\stackrel{\leftrightarrow}{\partial^{\nu}}\!\!\psil)$,
similar to that of the transverse part.
If we integrate Eq.\ (\ref{eq:ld1}) over $\tau$, one obtains a conservation
law of the same type as well.

Finally we state the equation for $\phil$, the four-spinor main field
\begin{equation}
\partial_{\tau}\rho_{\parallel}=\frac{i}{2M}\partial_{\parallel\mu}
(\phibarl\gdt\stackrel{\leftrightarrow}{\partial_{\parallel}^{\mu}}\phil)
\end{equation}
The solution of the equations of motion is given in Appendix
\ref{app:eqnsols}.

\section*{VI. Product Hilbert space}
\label{sec:prodhil}

The vector $t_{\mu}$ splits
the evolution of the momentum into two parts, the transverse and the
longitudinal.
These two modes of the motion of an event, transverse and longitudinal,
are complementary. Therefore, the overall Hilbert space is the tensor product
of the transverse and longitudinal Hilbert spaces
\begin{equation}
{\cal H}={\cal H}_{\perp}\otimes{\cal H}_{\parallel} \label{eq:Hilbert2}
\end{equation}
Considering the free event, the probability density is
\begin{equation}
\rho=\rho_{\perp}\cdot\rho_{\parallel}
\end{equation}
By defining
\begin{eqnarray}
L_{\perp}  & = & \gdt\sigmapt  \\
L_{\parallel} & = & i\gdt(P\cdot t) \\
\omega & = & 2M
\end{eqnarray}
and absorbing the signs $\splus,\sminus$ into $\phi$ or $\chi$,
the transverse and longitudinal equations can be expressed in a similar manner
\begin{eqnarray}
L_{\perp}\phi_{\perp} & = & \omega\chi_{\perp} \\
L_{\perp}\chi_{\perp} & = & \idtau\phi_{\perp}
\end{eqnarray}
and
\begin{eqnarray}
L_{\parallel}\phi_{\parallel} & = & \omega\chi_{\parallel} \\
L_{\parallel}\chi_{\parallel} & = & \idtau\phi_{\parallel}
\end{eqnarray}
In the product Hilbert space we define
\begin{eqnarray}
\phi & = & \phip\otimes\phil \\
\chi & = & \chip\otimes\chil
\end{eqnarray}
and $L_{\perp},L_{\parallel}$ operate on their respective factors.
Now
\begin{eqnarray}
L_{\perp}\phi & = & \omega(\chip\otimes\phil) \\
L_{\parallel}\phi & = & \omega(\phip\otimes\chil) \\
L_{\perp}\chi & = & (\idtau\phip)\otimes\chil) \\
L_{\parallel}\chi & = & \chip\otimes(\idtau\phil)
\end{eqnarray}
Denoting
\begin{equation}
L=L_{\perp}\otimes L_{\parallel}
\end{equation}
we obtain
\begin{eqnarray}
L\phi & = & \omega^{2}\chi \\
L\chi & = & (\idtau\phip)\otimes(\idtau\phil)  \\
\end{eqnarray}
and
\begin{equation}
L^{2}\phi=L^{2}\chi=(\omega\idtau\phip)\otimes(\omega\idtau\phil)
\end{equation}

We are interested in getting a unified equation for the transverse and
longitudinal
modes, conforming to the full \schstu equation. Therefore we consider the
combinations
\begin{eqnarray}
(L_{\perp}+L_{\parallel})\phi & = & \omega(\chip\otimes\phil+
\phip\otimes\chil) \\
(L_{\perp}+L_{\parallel})\chi & = & (\idtau\phip)\otimes\chil+\chip\otimes
(\idtau\phil)
\end{eqnarray}
Considering $(L_{\perp}+L_{\parallel})^{2}$ we find
\begin{eqnarray}
(L_{\perp}+L_{\parallel})^{2}\phi & = & \omega\idtau\phi + 2\omega^{2}\chi
\nonumber \\
& = & \omega\idtau\phi + 2L\phi
\end{eqnarray}
Since
\begin{equation}
(L_{\perp}+L_{\parallel})^{2}=L_{\perp}^{2}+L_{\parallel}^{2}+2L
\end{equation}
we obtain
\begin{equation}
(L_{\perp}^{2}+L_{\parallel}^{2})\phi=\omega\idtau\phi
\end{equation}
which is the desired full \schstu equation. Repeating the same steps for the
auxiliary field $\chi$, we find that it also obeys the full \schstu equation
\begin{equation}
(L_{\perp}^{2}+L_{\parallel}^{2})\chi=\omega\idtau\chi
\end{equation}

\section*{VII. Interacting charged equations}
\label{sec:charged}

To introduce the electromagnetic coupling, we define the electromagnetic field
appropriately on the manifold of the tensor product space i.e.,
\begin{equation}
a^{\alpha}(x,\tau)\equiv a^{\alpha}(x_{\perp},x_{\parallel},\tau)
\label{eq:defx}
\end{equation}
where $\alpha=0,1,2,3,\tau$. We may therefore define
\begin{eqnarray}
L_{\perp}^{\Pi}  & = & \gdt\sigpit  \\
L_{\parallel}^{\Pi} & = & i\gdt(\Pi\cdot t)
\end{eqnarray}
where
\begin{equation}
\Pi^{\mu}=P^{\mu}-ea^{\mu}
\end{equation}
Note that although $L_{\perp}$ and $L_{\parallel}$ act on the tensor product
space as $L_{\perp}\otimes{\bf 1}$ and
${\bf 1}\otimes L_{\parallel}$, this is no longer true, in general, for
$L_{\perp}^{\Pi}$ and $L_{\parallel}^{\Pi}$.
Let us start by defining equations containing fields of the form
$a_{\perp\mu}$
and $a_{\parallel\mu}$
\begin{equation}
\gdt[\sigmunu(P_{\mu}-ea_{\perp\mu}(x_{\perp},\tau))t_{\nu}]
(\phip(x_{\perp},\tau)\otimes{\bf 1}) =
\omega(\chip(x_{\perp},\tau)\otimes{\bf 1})
\end{equation}
\begin{equation}
\gdt[\sigmunu(P_{\mu}-ea_{\perp\mu}(x_{\perp},\tau))t_{\nu}]
(\chip(x_{\perp},\tau)\otimes{\bf 1}) =
[(\idtau+ea_{\perp\tau}(x_{\perp},\tau))\phip(x_{\perp},\tau)]\otimes{\bf 1}
\end{equation}
and
\begin{equation}
i\gdt[(P_{\mu}-ea_{\parallel\mu}(x_{\parallel},\tau))t^{\mu}]
({\bf 1}\otimes\phil(x_{\parallel},\tau)) =
\omega({\bf 1}\otimes\chil(x_{\perp},\tau))
\end{equation}
\begin{equation}
i\gdt[(P_{\mu}-ea_{\parallel\mu}(x_{\parallel},\tau))t^{\mu}]
({\bf 1}\otimes\chil(x_{\parallel},\tau)) =
{\bf 1}\otimes[(\idtau+ea{\parallel\tau}
(x_{\parallel},\tau))\phil(x_{\parallel},\tau)]
\end{equation}
Now, since we can write
\begin{equation}
\sum(a_{\perp}^{n}\phip\otimes a_{\parallel}^{n}\phil)
(x_{\perp},x_{\parallel})=\sum a_{\perp}^{n}(x_{\perp})a_{\parallel}^{n}
(x_{\parallel})\phip(x_{\perp})\phil(x_{\parallel})
\end{equation}
where $n$ is some index, the limit of this sum can approximate any function
$a(x)$ i.e.,
\begin{equation}
\sum a_{\perp}^{n}(x_{\perp})a_{\parallel}^{n}(x_{\parallel})
\phip(x_{\perp})\phil(x_{\parallel})=a(x)\phip(x_{\perp})\phil(x_{\parallel})
\end{equation}
Since we have gauge invariance
\begin{eqnarray}
a_{\perp\mu}(x_{\perp}) & \rightarrow & a_{\perp\mu}(x_{\perp})+
\frac{1}{e}\partial_{\perp\mu}\Lambda_{\perp}(x_{\perp}) \\
a_{\parallel\mu}(x_{\parallel}) & \rightarrow &
a_{\parallel\mu}(x_{\parallel})+\frac{1}{e}\partial_{\parallel\mu}
\Lambda_{\parallel}(x_{\parallel})
\end{eqnarray}
we may generalize the gauge invariance to
\begin{equation}
a_{\mu}(x_{\perp},x_{\parallel},\tau)\rightarrow a_{\mu}
(x_{\perp},x_{\parallel},\tau)+\frac{1}{e}\partial_{\mu}
\Lambda(x_{\perp},x_{\parallel},\tau)
\end{equation}
and
\begin{equation}
a_{\tau}(x_{\perp},x_{\parallel},\tau)\rightarrow
a_{\tau}(x_{\perp},x_{\parallel},\tau)-\frac{1}{e}
\partial_{\tau}\Lambda(x_{\perp},x_{\parallel},\tau)
\end{equation}

These hold for the equations
\begin{eqnarray}
\gdt(\sigma\Pi t)_{\perp}(\phip\otimes\phil) & = &
\omega(\chip\otimes\phil) \\
\gdt(\sigma\Pi t)_{\perp}(\chip\otimes\chil) & = &
(\idtau\phip)\otimes\chil+e\atau(\phip\otimes\chil) \\
i\gdt(\Pi\cdot t)_{\parallel}(\phip\otimes\phil) & = &
\omega(\phip\otimes\chil) \\
i\gdt(\Pi\cdot t)_{\parallel}(\chip\otimes\chil) & = &
\chip\otimes(\idtau\phil)+e\atau(\chip\otimes\phil)
\end{eqnarray}
since in each equation we can refer to the other variable
(say $\parallel$ in the $\perp$ equation), as a parameter.
The subscript on the matrix operators on the L.H.S. of the equations
means that we operate with derivatives on the relevant factor space.

We can therefore define the form of the basic equations as
\begin{eqnarray}
L_{\perp}^{\Pi}\phi_{\perp} & = & \omega\chi_{\perp} \\
L_{\perp}^{\Pi}\chi_{\perp} & = & (\idtau+e\atau)\phi_{\perp}
\end{eqnarray}
and
\begin{eqnarray}
L_{\parallel}^{\Pi}\phi_{\parallel} & = & \omega\chi_{\parallel} \\
L_{\parallel}^{\Pi}\chi_{\parallel} & = & (\idtau+e\atau)\phi_{\parallel}
\end{eqnarray}
Denoting $L^{\Pi}$ as the tensor product operator obtained from
$L_{\perp}^{\Pi}$ and $L_{\parallel}^{\Pi}$, as
\begin{equation}
L^{\Pi}=L_{\perp}^{\Pi}\otimes L_{\parallel}^{\Pi}
\end{equation}
we obtain
\begin{eqnarray}
L^{\Pi}\phi & = & \omega^{2}\chi \\
L^{\Pi}\chi & = & (\idtau\phip)\otimes(\idtau\phil)+e^{2}\atau^{2}\phi+
e\atau\idtau\phi
\end{eqnarray}
As before, we obtain from the squared expression
\begin{eqnarray}
(L_{\perp}^{\Pi}+L_{\parallel}^{\Pi})^{2}\phi & = &
\omega(\idtau+e\atau)\phi+2\omega^{2}\chi \nonumber \\
& = & \omega(\idtau+e\atau)\phi+2L^{\Pi}\phi
\end{eqnarray}
Therefore we find
\begin{equation}
[(L_{\perp}^{\Pi})^{2}+(L_{\parallel}^{\Pi})^{2}]\phi=
\omega(\idtau+e\atau)\phi \label {eq:fcsse1}
\end{equation}
which is exactly the full electromagnetically coupled \schstu equation.
Now, using the fact that
\begin{equation}
(\sigmunu \Pi_{\mu} t_{\nu})(\sigrolam \Pi_{\rho} t_{\lambda})=
\frac{1}{4}(\mbox{[}\Pi_{\mu},\Pi_{\rho}\mbox{]}\mbox{[}\sigmunu
t_{\nu},\sigmuro t_{\lambda}\mbox{]} +
\{\Pi_{\mu},\Pi_{\rho}\}\{\sigmunu t_{\nu},\sigmuro t_{\lambda}\}   )
\end{equation}
the definition (this quantity was defined by Horwitz and Arshansky \cite{HA82}
as well)
\begin{equation}
-2i\sigma_{\small{t}}^{\mu\rho}=[\sigmunu t_{\nu},\sigrolam t_{\lambda}]=
2i(\sigmuro t^{2}-(\sigmulam t_{\lambda})t^{\rho}-(\signuro t_{\nu})t^{\mu})
\end{equation}
and
\begin{eqnarray}
\{\sigmunu v_{\nu},\sigrolam v_{\lambda}\}=2(v^{2}\gmuro-v^{\mu}v^{\rho})
\label{eq:siganticom}
\end{eqnarray}
and the relation
\begin{equation}
[\Pi_{\mu},\Pi_{\nu}]=ief_{\mu\nu}
\end{equation}
we get
\begin{equation}
(L_{\perp}^{\Pi})^{2} = \Pi_{\perp}^{2}-\frac{e}{2}
\sigma_{t}^{\mu\nu}f_{\mu\nu}
\end{equation}
by defining
\begin{equation}
\Pi_{\perp\mu}\equiv\Pi_{\mu}+(\Pi\cdot t)t_{\mu}
\end{equation}
Furthermore, we have
\begin{eqnarray}
(L_{\parallel}^{\Pi})^{2} & = & -(\Pi\cdot t)^{2} \nonumber \\
\Pi_{\perp}^{2} & = & \Pi^{2}+(\Pi\cdot t)^{2} \nonumber
\end{eqnarray}
and it is easy to see that Eq.\ (\ref{eq:fcsse1}) is
\begin{equation}
(\idtau +e\atau)\phi =\frac{1}{2M}[\Pi^{2}-\frac{e}{2}
\sigma_{\small{t}}^{\mu\nu}f_{\mu\nu}]\phi
\end{equation}
which is exactly the same equation found by Horwitz and Arshansky \cite{HA82}.
This shows that our formulation is fully consistent with theirs.

The gyromagnetic ratio is taken as the ratio between
the coefficients of the terms
$\Pi^{2}$ and $\frac{e}{2}\sigma_{\small{t}}^{\mu\nu}f_{\mu\nu}$, giving
the correct relation of relative size and sign, between the momentum, spin
coupling, and mass terms (see \cite{ITZUB}).
Furthermore, we obtain, as in Horwitz and Arshansky \cite{HA82},
a fully Hermitian spin term, as opposed to the Dirac case (and the additional
$\atau$ field which they did not use).


\section*{VIII. The Dirac limit}
\label{sec:dirconn}

We have seen that it is impossible under the assumptions of Section
\ref{sec:genfeat}
to find a single first order equation which iterates to the full \schstu
equation. Therefore, we resorted to the combined product Hilbert space
description in Section \ref{sec:prodhil}.
However, by making a slight modification we can obtain the full
\schstu equation by iteration from a single equation, but there is a price to
pay; the gyromagnetic
term in the electromagnetically coupled version is no longer Hermitian,
like in Dirac's three dimensional case, and the structure of the continuity
equation is lost.
This modification is the necessary ingredient to see the connection between
the theory developed so far and Dirac's theory.

The modified approach that we use to employ a single equation is to break up
the operator $L(P)$ to two parts, which have different algebraic structure
with
respect to the nilpotents $N_{\pm}$. This will enable us to obtain the full
\schstu equation upon iteration.

Let us assume that $L(P)$ is composed explicitly of two parts, $L_{1}(P)$ and
$L_{2}(P)$. Then, the general form of the first order equation is
\begin{equation}
(L_{1}(P)+L_{2}(P))\psi=\sminus\nminus\idtau\psi +\splus M\nplus\psi
\label{eq:modform}
\end{equation}
We now multiply Eq.\ (\ref{eq:modform}) from the left by
$(L_{1}(P)-L_{2}(P))$. To be able to insert Eq.\ (\ref{eq:modform}) into the
new equation we must have
\begin{eqnarray}
\{ L_{1}(P),N_{\pm} \}=0 \nonumber \\
\mbox{[} L_{2}(P),N_{\pm} \mbox{]}=0
\end{eqnarray}
Requiring that
\begin{eqnarray}
\mbox{[} L_{1}(P),L_{2}(P) \mbox{]}=0
\end{eqnarray}
and that $N_{\pm}$ are nilpotents, we obtain
\begin{equation}
(L_{1}^{2}(P)-L_{2}^{2}(P))\psi=-\sminus\splus M\antiCnn\idtau\psi
\end{equation}
Performing an evaluation process similar to the one done in Section
\ref{sec:genfeat},
we make the simplest choice which is
\begin{eqnarray}
L_{1}(P)\equiv\sigmapt & ; & L_{2}(P)\equiv i(P\cdot t)
\end{eqnarray}
and the nilpotents are $N_{\pm}=\gdt({\bf 1}\pm\gamaf)$.
This is a synthesis of the use of commutators and anticommutators of Section
\ref{sec:genfeat}, and of the transverse and longitudinal equations.
The equation of motion is
\begin{equation}
(-\sigma^{\mu\nu}i\partial_{\mu}t_{\nu}+i(-i\dmu t^{\mu}))\psi=
\sminus\gdt\pminus\idtau\psi+\splus\frac{M}{2}\gdt\pplus\psi
\label{eq:jointeq}
\end{equation}
Iterating we get the {\sl full} \schstu equation (\ref{eq:sse}).
We obtain the currents
\begin{equation}
-i\dnu(\psibar\sigma^{\nu\mu}t_{\mu}\psi)+\psibar\!
\stackrel{\leftrightarrow}{\partial^{\mu}}\!\!t_{\mu}\psi)=
2\sminus\idtau(\psibar_{L}\gdt\psi_{L}) \label{eq:L1}
\end{equation}
which can be written as
\begin{equation}
-t_{\mu}j_{L}^{\mu}=2\dtau\sminus t_{\mu}(\psibar_{L}\gamma^{\mu}\psi_{L})
\label{eq:gdc}
\end{equation}
where
\begin{equation}
j_{L}^{\mu}\equiv(\dnu(\psibar\sigma^{\nu\mu}\psi)+\psibar i\!
\stackrel{\leftrightarrow}{\partial^{\mu}}\!\!\psi)
\end{equation}
It is clearly seen that $j_{L}^{\mu}$ resembles the definition of the
Gordon decomposition of currents of the Dirac equation.

An equation with the roles of $N_{\pm}$ interchanged can built, resulting in
a similar Gordon decomposition but this time only right-right terms appear
in the current. When we discussed the transverse and longitudinal
equations we could transform between the two versions opposite in the
nilpotent assignment by a suitable transformation.
Decomposing Eq.\ (\ref{eq:jointeq}) by projection into two coupled equations
we find
\begin{eqnarray}
-i(\gamma\cdot P)\psiL & = & \splus M\psiR \nonumber \\
-i(\gamma\cdot P)\psiR & = & 2\sminus\idtau\psiL \label{eq:tindep}
\end{eqnarray}
At this stage the $t_{\mu}$ dependence of the equations disappears; we deal
with this below.
Taking these equations to mass shell, choosing $\splus=\sminus=-1$,
and observing the eigenvalues of plane wave solutions (see Appendix
\ref{app:eqnsols}),
we obtain
\begin{eqnarray}
-i(\gamma\cdot P)\psiL & = & -m\psiR \nonumber \\
-i(\gamma\cdot P)\psiR & = & m\psiL \label{eq:psidir}
\end{eqnarray}
and for the equation with $N_{\pm}$ interchanged
\begin{eqnarray}
-i(\gamma\cdot P)\varphiR & = & -m\varphiL \nonumber \\
-i(\gamma\cdot P)\varphiL & = & m\varphiR \label{eq:varphidir}
\end{eqnarray}
To transform from Eq.\ (\ref{eq:psidir}) to Eq.\ (\ref{eq:varphidir}), we can
multiply $\psi$ by $-\frac{1}{m}(\gamma\cdot P)$ (working on shell) to
exchange the roles of the left and right handed spinors,
therefore the relation between $\psi$ and $\varphi$ is
\begin{eqnarray}
-\frac{\sigma^{\mu}p_{\mu}}{m}\psiR=\varphiL & ; &
-\frac{\tilde{\sigma}^{\mu}p_{\mu}}{m}\psiL=\varphiR \label{eq:mydirsols}
\end{eqnarray}
This implies that
\begin{eqnarray}
\varphiL=-i\psiL & ; & \varphiR=i\psiR \label{eq:direl}
\end{eqnarray}
and the relation between the main field $\phi$ and the auxiliary field
$\chi$ is
\begin{equation}
\phi=i\chi \label{eq:maintoaux}
\end{equation}
Using Eq.\ (\ref{eq:direl}), Eqs.\ (\ref{eq:psidir}),(\ref{eq:varphidir}) are
then exactly the Dirac equation in the chiral representation
(holding for both the main and auxiliary fields)
\begin{eqnarray}
(\gamma\cdot P)\phi_{R} & = & -m\phi_{L} \nonumber \\
(\gamma\cdot P)\phi_{L} & = & -m\phi_{R}
\end{eqnarray}
This implies, of course, the Dirac current and continuity equations for the
main and auxiliary fields $\phi$ and $\chi$.

When electromagnetic coupling is introduced in Eq.\ (\ref{eq:jointeq}),
assuming that $\atau=0$ (Coulomb-like gauge)
and $a_{\mu}$ independent of $\tau$ (restricting ourselves to the zero
mode) for simplicity, we obtain for the second order equation
\begin{equation}
\idtau\psi =\frac{1}{2M}[\Pi^{2}-\frac{e}{2}\sigma^{\mu\nu}f_{\mu\nu}]\psi
\end{equation}
which has the usual non-hermitian term, and conforms to the second order
charged Dirac equation (when $\idtau\rightarrow -\frac{m^{2}}{2M}$, see
\cite{Schw50}).
We therefore see that the decomposition into transverse and longitudinal modes
was essential to achieve a Hermitian interaction as well as the \schstu form.

All the discussion above can be approached from a different point of view.
If we
take the longitudinal equations of motion, and assume that the momentum is in
the direction of the four-vector $t_{\mu}$, such that
\begin{eqnarray}
P_{\mu}=\alpha t_{\mu} & ; & P^{2}=-\alpha^{2}
\end{eqnarray}
then the longitudinal equations of motion become
\begin{eqnarray}
-i(\gamma\cdot P)\psiLl & = & \splus M\psiRl \nonumber \\
-i(\gamma\cdot P)\psiRl & = & 2\sminus\idtau\psiLl
\end{eqnarray}
with a similar result for the nilpotent interchanged equation.
Making the identification $\psi=\psil$, these are exactly
Eqs.\ (\ref{eq:tindep}).
One can now understand how $t_{\mu}$ disappeared previously in
Eqs.\ (\ref{eq:tindep}).
Furthermore, looking at the basic structure of the solutions of the
longitudinal equation for $\phil$ (see Appendix \ref{app:eqnsols}),
Eq.\ (\ref{eq:longsols1}),
and replacing $t_{\mu}$ by $\frac{p_{\mu}}{m}$ in $\zeta_{\parallel}^{-}(t)$,
(on shell condition), we obtain the exact non-normalized solutions of the
Dirac equation in the chiral representation
\begin{eqnarray}
\zeta_{1}(p) \equiv \left( \begin{array}{c}
-\frac{1}{m}\sigma^{\mu} p_{\mu}\xi_{1} \\
\xi_{1}
\end{array} \right)
&
\zeta_{2}(p) \equiv \left( \begin{array}{c}
\xi_{2} \\
-\frac{1}{m}\tilde{\sigma}^{\mu}p_{\mu}\xi_{2}
\end{array} \right)
\label{eq:diracsols}
\end{eqnarray}
where $\xi_{1,2}$ are two independent two-spinors; this is also apparent from
Eq.\ (\ref{eq:mydirsols}) (actually, when taken in the energy representation
of the Dirac matrices these solutions correspond to the positive energy,
and $\zeta_{\parallel}^{+}(t)$ solves the Dirac equation for the negative
energy). The transformation of the
extended helicity projection operator (see Appendix \ref{app:eqnsols}) is a
bit delicate since $\sigmapt$ and
$\mid p_{\perp}\mid$ go to zero together. Observing Eq.\ (\ref{eq:jointeq})
and its consequent Eqs.\ (\ref{eq:tindep}), we find the extended helicity
operator commutes with them. In Eqs.\ (\ref{eq:tindep}) $t_{\mu}$ does not
appear, so we can take it to be $t_{\mu}=(1,0,0,0)$, and the extended helicity
operator becomes the usual helicity.
We use the usual helicity operator to obtain four distinct solutions.

On the other hand, for the transverse equation, taking $p_{\mu}$ in the
direction of $t_{\mu}$ has different consequences. Observing
Eq.\ (\ref{eq:eqtn1}),
\begin{equation}
(\sigma^{\mu\nu}P_{\mu}P_{\nu})\psi_{\perp}=\sminus(\gamma\cdot P)\pminus
\idtau\psi_{\perp}+\splus\frac{M}{2}(\gamma\cdot P)\pplus\psi_{\perp}
\end{equation}
then $\sigma^{\mu\nu}P_{\mu}P_{\nu}=0$,
and $p_{\perp}^{2}=0$. Considering the free solutions, there is no $\tau$
evolution,
\begin{equation}
0=\splus\frac{M}{2}(\gamma\cdot P)\pplus\psi_{\perp}
\end{equation}
and we can say that
$\psiRp$ and $\varphiLp$ are zero, there is no auxiliary field, and no
transverse first order equations exist. All the evolution is described by the
longitudinal equation.

It seems that as $p_{\mu}$ departs from the direction of $t_{\mu}$, we depart
from the Dirac description, and the event is described by two independent
equations of motion, transverse and longitudinal. The event evolution in
$\tau$ becomes different for the two versions, differing in phase and
spinor content. At the same time the auxiliary fields
are spontaneously generated in the transverse case, and from the Dirac
field in the longitudinal case.
Furthermore, the auxiliary field, treated as a mathematical convenience,
but seen to be strongly related to the main field, Eq.\ (\ref{eq:maintoaux}),
may be a further indication of the amount of departure from the on shell
Dirac theory.


\section*{IX. Conclusions}
\label{sec:conclusions}

Accepting a second order spin equation of the form of the \schstu equation
as the basic structure of the theory, and requiring that the free solutions of
a first order equation be also the solutions of the second order one, lead
to transverse and longitudinal equations. After a selection
process we chose the most suitable forms for the ingredients of these
equations. We introduced the induced representations on a time-like vector
$t_{\mu}$ and an auxiliary field so that the theory be consistent. The
evolution
of the free event is governed by two complementary equations, the transverse
and longitudinal. We manage to unify the complementary behavior of the event,
transverse and longitudinal, with the use of a product Hilbert space.
Then, after introducing the electromagnetic coupling,
we obtain the correct gyromagnetic ratio, and the second order
charged equation describing the evolution of the main field, is fully
Hermitian under the scalar product.
Finally, we showed how the theory goes over to the Dirac theory in the limit
in which $p_{\mu}$
is in the direction of $t_{\mu}$ on mass shell. In this limit the theory
changes its structure.
The first order transverse equation no longer exists; its
auxiliary field and $\tau$ evolution disappear, and the longitudinal equation
takes on a new meaning, leading to the original Dirac equation.
We also display, in Appendix \ref{app:eqnsols}, the solutions of the
equations, which are characterized
by the ``extended helicity'' and ``chiral precedence'' for the transverse
case, and
by ``extended helicity'' and ``extended parity'' in the longitudinal case.
Both cases have the same properties under the generalized parity, charge
conjugation,
and $\tau$ reversal transformations (as shown in Appendix \ref{app:pct}). The
objects which transform one into the
other under these transformations are pure left or right handed spinors, thus
exhibiting the chiral nature of the theory.

\appendix
\newpage


\section{Idempotents and Nilpotents}
\label{app:all_potents}


\subsection{Idempotents}
\label{appsubs:idem}

Let us consider idempotents of the form (we carry out the analysis
in a given Lorentz frame, and discuss later the corresponding invariant
forms)
\begin{eqnarray}
P_{i\pm} & = & \frac{1}{2}({\bf 1}\pm\Gamai) \nonumber \\
\end{eqnarray}
where $\Gamai$ are elements of the Dirac Clifford algebra
(see Appendix \ref{app:diracmat}).
We state some known facts about these matrices \cite{Thal}
\begin{eqnarray}
\Gamai^{2} & = & {\bf 1}   \label{eq:Gamaone} \\
\Gamai^{-1} & = & \Gamai   \label{eq:Gamatwo} \\
\Gamai^{\dagger} & = & \Gamai \;\;\;\;\;\;\;\;\;\;\;\; (i=1,\ldots,16)
\end{eqnarray}
where ``$^{\dagger}$'' implies complex conjugate transpose, and for $j\neq k$,
\begin{eqnarray}
\Gamaj\Gamak & =& \epsilon_{jk}\Gamal \;\;\;\;\;\;\;\;
\epsilon_{jk}\in\{1,-1,i,-i\}  \label{eq:eps}\\
\Gamak\Gamaj & = & (\epsilon_{jk})^{-1}\Gamal   \label{eq:ieps}
\end{eqnarray}
There are fifteen pairs of $P_{i\pm}$.  In fact
they are equivalent under automorphism. We show this by defining a unitary
transformation $\Tj$ such that
\begin{eqnarray}
\Tj =\frac{1}{\sqrt{2}}({\bf 1}+i\Gamaj) \;\; ; \;\; \Tj^{-1}=
\frac{1}{\sqrt{2}}({\bf 1}-i\Gamaj)=\Tj^{\dagger}
\end{eqnarray}
Using $\Tj$ we can transform from one $\Gamma$ to another. If $\Gamai$
and $\Gamaj$ commute we have
\begin{equation}
[\Gamaj,\Gamai]=0  \Rightarrow  [\Tj,\Gamai]=0  \Rightarrow
\Tj\Gamai\Tj^{-1}=\Gamai
\end{equation}
and because of Eqs.\ (\ref{eq:eps}),(\ref{eq:ieps}) we have (for $i\neq j$)
\begin{eqnarray}
[\Gamaj,\Gamai]=0  & \Rightarrow & \Gamai\Gamaj-\Gamaj\Gamai=
(\epsilon_{ij}-\epsilon_{ij}^{-1})\Gamak =0 \nonumber \\
 & \Rightarrow & \epsilon_{ij}=\epsilon_{ij}^{-1}  \Rightarrow
 \epsilon_{ij}=\pm 1
\end{eqnarray}
and
\begin{eqnarray}
[\Gamai,\Gamaj]=0 & \Rightarrow & \{\Gamai,\Gamaj\}\neq 0
\end{eqnarray}
$\Gamai$ is invariant under the transformation $\Tj$; we find
a triplet $\Gamai$, $\Gamaj$, $\Gamak$ which commute among
themselves, where $\Gamak=\Gamai\Gamaj$. These triplets form maximal
commuting sets.

On the other hand, if $\Gamai$ and $\Gamaj$ do not commute then
\begin{eqnarray}
[\Gamaj,\Gamai]\neq 0  & \Rightarrow & [\Tj,\Gamai]\neq 0 \Rightarrow
(\epsilon_{ij}-\epsilon_{ij}^{-1})\Gamak\neq 0 \nonumber \\
 & \Rightarrow  & \epsilon_{ij}\neq\epsilon_{ij}^{-1} \Rightarrow
 \epsilon_{ij}=\pm i
\end{eqnarray}
This means that for the non-commuting case we have
\begin{equation}
[\Gamaj,\Gamai]=\pm 2i\Gamak
\end{equation}
and so
\begin{eqnarray}
\Tj\Gamai\Tj^{-1} & = & \frac{1}{2}
(\Gamai +i[\Gamaj,\Gamai]+\Gamaj\Gamai\Gamaj) = \mp\Gamak
\end{eqnarray}
Furthermore we find that
\begin{eqnarray}
[\Gamai,\Gamaj]\neq 0   & \Rightarrow & \{\Gamai,\Gamaj\}=0
\end{eqnarray}
We can transform $\Gamai$ into $\Gamak$ using $\Gamaj$.
This is a Pauli algebra structure for the triplet $\Gamai$, $\Gamaj$,
$\Gamak$.
\begin{eqnarray}
[\Gamaj,\Gamai]=\pm 2i\Gamak & ; & \{\Gamaj,\Gamai\}=0 \nonumber \\
\mbox{[}\Gamai,\Gamak]=\pm 2i\Gamaj & ; & \{\Gamai,\Gamak\}=0 \nonumber \\
\mbox{[}\Gamak,\Gamaj]=\pm 2i\Gamai & ; & \{\Gamak,\Gamaj\}=0
\end{eqnarray}


From $P_{i\pm}$ and $P_{j\pm}$ from a maximal commuting set, we can form four
primitive idempotents
\begin{eqnarray}
P_{i\pm}P_{j\pm} & = & (\frac{1}{2})^{2}({\bf 1}\pm\Gamai)
({\bf 1}\pm\Gamaj)  \\
\end{eqnarray}
We shall denote these by
\begin{eqnarray}
\tilde{P}_{\alpha}=P_{i\pm}P_{j\pm} & (\alpha =1,2,3,4)
\end{eqnarray}
where
\begin{displaymath}
\begin{array}{lllll}
\alpha & = & 1 & \longrightarrow & i+,j+ \\
\alpha & = & 2 & \longrightarrow & i+,j- \\
\alpha & = & 3 & \longrightarrow & i-,j+ \\
\alpha & = & 4 & \longrightarrow & i-,j-
\end{array}
\end{displaymath}
The triplets of nontrivial commuting $\Gamma$'s are
(the numbers refer to the index of the matrices, see Appendix
\ref{app:diracmat}):
\begin{displaymath}
\begin{array}{rrrrr}
 6,9,16  &  6,4,15  &  6,5,14  &  9,2,13  &  9,3,12 \\
16,8,11  & 16,7,10  & 15,2,11  &  15,3,7  & 10,4,12 \\
2,10,14  & 11,5,12  &  14,3,8  &  13,5,7  &  8,4,13
\end{array}
\end{displaymath}
For building $\tilde{P}_{\alpha}$'s we can take any pair from a given
triplet, and
achieve the same result (get the same $\tilde{P}_{\alpha}$'s for the
triplet); this is because
\begin{eqnarray}
\tilde{P}_{\alpha} & = & (\frac{1}{2})^{2}({\bf 1}\pm\Gamai\pm\Gamaj\pm
\Gamai\Gamaj)
\end{eqnarray}
and
\begin{equation}
\Gamai\Gamaj =\pm\Gamak
\end{equation}
Furthermore, each $\Gamma$
appears only in {\sl three} of the above triplets, and we can transform from
one triplet to another in this ``trio'' of triplets, by a unitary
transformation,
thus transforming between different sets of $\tilde{P}_{\alpha}$'s.
If we denote
the trio of triplets in which a $\Gamma$ matrix (say, ``a'') appears as
\begin{displaymath}
\begin{array}{l}
\fbox{a},b,c  \\
\fbox{a},d,e  \\
\fbox{a},f,g
\end{array}
\end{displaymath}
then we find that we always have a Pauli type relation of anticommutators
among the remaining pairs, for example such as:
\begin{displaymath}
\begin{array}{ccccccc}
\{b,d\} & = & \{b,f\} & = & \{d,f\} & = & 0  \\
\{b,e\} & = & \{b,g\} & = & \{e,g\} & = & 0  \\
\end{array}
\end{displaymath}
here we took ``b'' as the ``pivot'' (i.e. ``b'' is used for $T_{b}$).
This enables us to take $T_{b}$ and transform the second triplet into the
third: $a,d,e \rightarrow a,f,g$ (it can be done equally well with ``c'' as
a ``pivot''), thus getting a transformed set of $\tilde{P}_{\alpha}$'s.
This way we can take a triplet in a trio, and use it to transform between
the other two triplets. Looking at the existing triplets we see that we can
transform from any triplet to any other by a suitable choice of unitary
transformations, i.e., all sets of $\tilde{P}_{\alpha}$'s are equivalent
up to a unitary transformation.


\subsection{Nilpotents}
\label{appsubs:nil}

We can produce non-primitive level nilpotents by taking
\begin{eqnarray}
n_{ij} & = & (\Gamai\pm i\Gamaj) \nonumber \\
n^{2}_{ij} & = & \pm i\{\Gamai,\Gamaj\}
\end{eqnarray}
and requiring that $\{\Gamai,\Gamaj\}=0$.

Now we turn to primitive level nilpotents.
In a certain representation we have $\tilde{P}_{k}=e_{kk}, (k=1,2,3,4)$
where $e_{kk}$ is
the matrix which has a ``1'' in the $k^{th}$ row and $k^{th}$ column, and
all other
places are zero. Since the matrices $e_{nm}, (n\neq m)$ are nilpotent, (we
have twelve of these), we get for any Dirac matrix $A$:
\begin{equation}
e_{ii}A e_{kk}=\alpha_{ik}e_{ik}
\end{equation}
where $\alpha_{ik}$ is the element on the $i^{th}$ row, $k^{th}$ column,
in $A$. Once
we have these twelve nilpotents, others can be built from them. If we
want to represent the nilpotents with the help of the $\Gamma$ matrices we
need to have
\begin{eqnarray}
N_{ikj} & = & \tilde{P}_{i}\Gamak\tilde{P}_{j} \nonumber \\
N^{2}_{ikj} & = & \tilde{P}_{i}\Gamak\tilde{P}_{j}\tilde{P}_{i}\Gamak
\tilde{P}_{j}=0 \;\;\;\;\; (i\neq j)
\end{eqnarray}
this is because $\tilde{P}_{j}\tilde{P}_{i}=0$ always. We must guard against
$N_{ikj}$ being zero, therefore we {\sl must not} have
$[\tilde{P}_{i},\Gamak]=0$ or $[\tilde{P}_{j},\Gamak]=0$ and $i\neq\ j$.
In this case, for $i=j$, the representation is diagonal.
Actually if we take a closer
look at $\tilde{P}_{i},\tilde{P}_{j}$, for example,
\begin{eqnarray}
\tilde{P}_{i}=P_{n\pm}P_{m\pm} & \tilde{P}_{j}=P_{n\pm}P_{m\mp}
\end{eqnarray}
then if $\{\Gamak,\Gamma_{m}\}=0$, $\Gamak P_{m\pm}=P_{m\mp}\Gamak$, and
$P_{m}$ flips sign. We can summarize the nonzero useful nilpotents as follows:
\begin{enumerate}
\item   \begin{eqnarray}
\{\Gamak,\Gamma_{m}\} = 0  & ; &  \mbox{[}\Gamak,\Gamma_{n}\mbox{]}=0
\;\;\;\; \Longrightarrow  \nonumber
\end{eqnarray}
\begin{eqnarray}
\tilde{P}_{i}\Gamak\tilde{P}_{j} & = & P_{n\pm}P_{m\pm}\Gamak P_{n\pm}
P_{m\mp}  =  \Gamak P_{n\pm}P_{m\mp}  =  \Gamak\tilde{P}_{j} \nonumber \\
\tilde{P}_{i^{'}}\Gamak\tilde{P}_{j^{'}} & = & P_{n\mp}P_{m\pm}\Gamak
P_{n\mp}P_{m\mp}  =  \Gamak P_{n\mp}P_{m\mp}  =  \Gamak\tilde{P}_{j^{'}}
\end{eqnarray}
\item     \begin{eqnarray}
\{\Gamak,\Gamma_{m}\}=0 & ; & \{\Gamak,\Gamma_{n}\}=0 \;\;\;\; \Longrightarrow
\nonumber
\end{eqnarray}
\begin{eqnarray}
\tilde{P}_{i}\Gamak\tilde{P}_{j} & = & P_{n\pm}P_{m\pm}\Gamak P_{n\mp}
P_{m\mp}  =  \Gamak P_{n\mp}P_{m\mp}  =  \Gamak\tilde{P}_{j} \nonumber \\
\tilde{P}_{i^{'}}\Gamak\tilde{P}_{j^{'}} & = & P_{n\pm}P_{m\mp}\Gamak
P_{n\mp}P_{m\pm}  =  \Gamak P_{n\mp}P_{m\pm}  =  \Gamak\tilde{P}_{j^{'}}
\end{eqnarray}
\end{enumerate}
thus showing that there are $N_{ikj}\neq 0$.
The first entry results in four nilpotents which are equivalent to the four
generated by the second entry up to a sign. Since both cases are equivalent
we shall concentrate on the second case.
We find that any
two commuting $\Gamma$'s, have {\sl exactly} four mutual anticommuting
$\Gamma$'s in common. Taking the two commuting $\Gamma$'s as a pair from a
triplet, then for each pair the four anticommuting $\Gamma$'s are distinct.
Therefore we get a ``structure'' which can be illustrated by a specific
example; for the triplet $P_{2},P_{11},P_{15}$ the relevant structure is :
\begin{displaymath}
\begin{array}{lll}
P_{11},P_{15} & : & \Gamma_{9},\Gamma_{10},\Gamma_{13},\Gamma_{14} \\
P_{2},P_{11}  & : & \Gamma_{3},\Gamma_{4},\Gamma_{6},\Gamma_{7} \\
P_{15},P_{2}  & : & \Gamma_{5},\Gamma_{8},\Gamma_{12},\Gamma_{16}
\end{array}
\end{displaymath}
The $P$'s can be thought of as idempotent generators , and the $\Gamma$'s as
nilpotent generators.
The structure covers all $\Gamma$'s. From the two $P$'s in a row we
can build the four $\tilde{P}_{\alpha}$'s. We may take any $\Gamma$ from that
specific row to produce the four nilpotents (all $\Gamma$'s in a row produce
the same nilpotents). The three rows produce four distinct nilpotents each,
summing up to the basic twelve required for $4\times 4$ matrices.
The important thing to remember is that
by a unitary transformation we can transform from one triplet to another, and
thus from one structure to another, hence all structures are equivalent up
to a unitary transformation. A special feature of the above structure
($P_{2},P_{11},P_{15}$) is that it produces the idempotents and nilpotents
characterized by the matrices $e_{ij}$.


\subsection{Nilpotents for $N_{\pm}$}
\label{appsubs:nilsfor}

Since we are looking for $\nminus$ which gives us a positive definite
probability density, Eq.\ (\ref{eq:speccont}), we need $\nminus$ to include an
``anchor'' in the form of $\gamaz$, which assures a positive part, and up
to one more term (which should
be smaller or equal to the anchor since $\gamma$ matrices are orthogonal).
Now, considering the form of the currents in Eq.\ (\ref{eq:speccont}),
$\gamaz =\Gamma_{2}$ is the only anchor possible because of the appearance
of $\gamaz$ in $\psibar$, so we have to look at the row in the structure
where $\Gamma_{2}$ resides and calculate the nilpotents (all other rows will
not produce nilpotents with this anchor). We take for example the structure:
\begin{displaymath}
\begin{array}{lll}
P_{6},P_{16} & : & \Gamma_{2},\Gamma_{3},\Gamma_{12},\Gamma_{13} \\
P_{6},P_{9}  & : & \Gamma_{7},\Gamma_{8},\Gamma_{10},\Gamma_{11} \\
P_{9},P_{16} & : & \Gamma_{4},\Gamma_{5},\Gamma_{14},\Gamma_{15}
\end{array}
\end{displaymath}
and focus on the first row. Now we generate nilpotents $N_{i}, (i=1,2,3,4)$
\begin{eqnarray}
N_{1}  =   \Gamma_{2} P_{6-}P_{16+}  =  (\frac{1}{2})^{2}
   (\Gamma_{2}+i\Gamma_{3}+i\Gamma_{12}-\Gamma_{13}) \nonumber \\
N_{2}  =   \Gamma_{2} P_{6-}P_{16-}  =  (\frac{1}{2})^{2}
(\Gamma_{2}+i\Gamma_{3}-i\Gamma_{12}+\Gamma_{13}) \nonumber \\
N_{3}  =   \Gamma_{2} P_{6+}P_{16+}  =  (\frac{1}{2})^{2}
(\Gamma_{2}-i\Gamma_{3}+i\Gamma_{12}+\Gamma_{13}) \nonumber \\
N_{4}  =   \Gamma_{2} P_{6+}P_{16-}  =  (\frac{1}{2})^{2}
(\Gamma_{2}-i\Gamma_{3}-i\Gamma_{12}-\Gamma_{13}) \nonumber
\end{eqnarray}
In terms of the $\gamma$ matrices we get
\begin{eqnarray}
N_{1} & = & (\frac{1}{2})^{2}(\gamaz -\gamaone +\gamaz\gamaf+\gamaone\gamaf)
\nonumber \nonumber \\
N_{2} & = & (\frac{1}{2})^{2}(\gamaz -\gamaone -\gamaz\gamaf-\gamaone\gamaf)
\nonumber \nonumber \\
N_{3} & = & (\frac{1}{2})^{2}(\gamaz +\gamaone +\gamaz\gamaf-\gamaone\gamaf)
\nonumber \nonumber \\
N_{4} & = & (\frac{1}{2})^{2}(\gamaz +\gamaone-\gamaz\gamaf+\gamaone\gamaf)
\label{eq:fournils}
\end{eqnarray}
We consider these results as
given in a specific reference frame. The Lorentz invariant form of
these nilpotents is:
\begin{eqnarray}
N^{\prime}_{1} & = & (\gamma\cdot l^{(1-)})+(\gamma\cdot l^{(1+)})\gamaf
\nonumber \\
N^{\prime}_{2} & = & (\gamma\cdot l^{(1-)})-(\gamma\cdot l^{(1+)})\gamaf
\nonumber \\
N^{\prime}_{3} & = & (\gamma\cdot l^{(1+)})+(\gamma\cdot l^{(1-)})\gamaf
\nonumber \\
N^{\prime}_{4} & = & (\gamma\cdot l^{(1+)})-(\gamma\cdot l^{(1-)})\gamaf
\end{eqnarray}
where we used for the four-vectors $l^{(1\pm)}$ the notation
\begin{eqnarray}
l^{(1-)} & = & (l_{0},-l_{0},0,0) \nonumber \\
l^{(1+)} & = & (l_{0},l_{0},0,0)
\end{eqnarray}
and $l_{0}$ is some number.

The requirement for a positive definite probability density implies a
reduction
in half of the participating matrices in each nilpotent combination of
Eqs.\ (\ref{eq:fournils}); we can do this by reverting to the non-primitive
level nilpotents:
\begin{equation}
\begin{array}{cclcl}
N_{1}+N_{2} & = & \frac{1}{2}(\gamaz -\gamaone)             &
\rightarrow & (\gamma\cdot l^{(1-)}) \\ \\
N_{1}-N_{2} & = & \frac{1}{2}(\gamaz\gamaf +\gamaone\gamaf) &
\rightarrow & (\gamma\cdot l^{(1+)})\gamaf \\ \\
N_{1}+N_{3} & = & \frac{1}{2}(\gamaz +\gamaz\gamaf)         &
\rightarrow & \gamadt\pplus \\ \\
N_{1}-N_{3} & = & \frac{1}{2}(-\gamaone +\gamaone\gamaf)    &
\rightarrow & -(\gamma\cdot s^{(1)})\pminus  \\ \\
N_{3}+N_{4} & = & \frac{1}{2}(\gamaz +\gamaone)             &
\rightarrow & (\gamma\cdot l^{(1+)}) \\ \\
N_{3}-N_{4} & = & \frac{1}{2}(\gamaz\gamaf -\gamaone\gamaf) &
\rightarrow & (\gamma\cdot l^{(1-)})\gamaf \\ \\
N_{2}+N_{4} & = & \frac{1}{2}(\gamaz -\gamaz\gamaf)         &
\rightarrow & \gamadt\pminus \\ \\
N_{2}-N_{4} & = & \frac{1}{2}(-\gamaone -\gamaone\gamaf)    &
\rightarrow & -(\gamma\cdot s^{(1)})\pplus
\end{array}
\label{eq:eightnils}
\end{equation}
where the four-vectors $s^{(1)}$, and $t$ have been used
\begin{eqnarray}
t & = & (t_{0},0,0,0)  \nonumber     \\
s^{(1)} & = & (0,s_{1},0,0)
\end{eqnarray}
and $t_{0}, s_{1}$ are numbers. Going to non-primitive level nilpotents, we
get twice as many nilpotents (eight), which come in pairs differing in sign.
Each pair can be a good candidate for $\nplus$, $\nminus$. From the
nilpotents in Eqs.\ (\ref{eq:eightnils}), only $(\gamma\cdot l^{(1\pm)})$ and
$\gamadt({\bf 1}\pm\gamaf)$
are valid for a positive definite probability density (having $\gamadt$ as a
Lorentz covariant positive definite ``anchor''). Furthermore, it seems that
$(\gamma\cdot l^{(1\pm)})$
and $\gamadt({\bf 1}\pm\gamaf)$ should not be treated equivalently, since
they are derived from the same row in the same structure.

Carrying out the procedure of finding nilpotents on the twelve available
structures, (there are fifteen triplets, but three of them include
$\Gamma_{2}$
as an idempotent generator instead of a nilpotent generator), we find the
following nilpotents. We mention only the relevant row in the structure
containing $\Gamma_{2}$, and giving rise to nilpotents which have a chance
of being positive definite. The generating idempotents are in brackets
\begin{eqnarray}
\begin{array}{lclcll}
P_{9},(P_{6},P_{16})  & : & \Gamma_{2}, \Gamma_{3}, \Gamma_{12}, \Gamma_{13}
& \rightarrow & (\gamma\cdot l^{(1\pm)}) , &  \gamadt({\bf 1}\pm\gamaf) \\ \\
P_{15},(P_{4},P_{6})  & : & \Gamma_{2}, \Gamma_{3}, \Gamma_{7},  \Gamma_{11}
& \rightarrow & (\gamma\cdot l^{(1\pm)}),  & \gamadt\pm\sigmats^{(2)}   \\ \\
P_{14},(P_{5},P_{6})  & : & \Gamma_{2}, \Gamma_{3}, \Gamma_{8},  \Gamma_{16}
& \rightarrow & (\gamma\cdot l^{(1\pm)}),  & \gamadt\pm\sigmats^{(3)}   \\ \\
P_{9},(P_{3},P_{12})  & : & \Gamma_{2}, \Gamma_{6}, \Gamma_{13}, \Gamma_{16}
& \rightarrow & \gamadt\pm\sigmats^{(1)},  & \gamadt\pm i\gamaf \\ \\
P_{11},(P_{8},P_{16}) & : & \Gamma_{2}, \Gamma_{5}, \Gamma_{12}, \Gamma_{15}
& \rightarrow & (\gamma\cdot l^{(3\pm)}),  & \gamadt({\bf 1}\pm\gamaf) \\ \\
P_{10},(P_{7},P_{16}) & : & \Gamma_{2}, \Gamma_{4}, \Gamma_{12}, \Gamma_{14}
& \rightarrow & (\gamma\cdot l^{(2\pm)}),  & \gamadt({\bf 1}\pm\gamaf) \\ \\
P_{15},(P_{3},P_{7})  & : & \Gamma_{2}, \Gamma_{4}, \Gamma_{6},  \Gamma_{11}
& \rightarrow & \gamadt\pm\sigmats^{(1)},  & (\gamma\cdot l^{(2\pm)})   \\ \\
P_{10},(P_{4},P_{12}) & : & \Gamma_{2}, \Gamma_{7}, \Gamma_{14}, \Gamma_{16}
& \rightarrow & \gamadt\pm\sigmats^{(2)},  & \gamadt\pm i\gamaf \\ \\
P_{11},(P_{5},P_{12}) & : & \Gamma_{2}, \Gamma_{8}, \Gamma_{15}, \Gamma_{16}
& \rightarrow & \gamadt\pm\sigmats^{(3)},  & \gamadt\pm i\gamaf \\ \\
P_{14},(P_{3},P_{8})  & : & \Gamma_{2}, \Gamma_{5}, \Gamma_{6},  \Gamma_{10}
& \rightarrow & (\gamma\cdot l^{(3\pm)}),  & \gamadt\pm\sigmats^{(1)}   \\ \\
P_{13},(P_{5},P_{7})  & : & \Gamma_{2}, \Gamma_{4}, \Gamma_{8},  \Gamma_{9}
& \rightarrow & (\gamma\cdot l^{(2\pm)}),  & \gamadt\pm\sigmats^{(3)}   \\  \\
P_{13},(P_{4},P_{8})  & : & \Gamma_{2}, \Gamma_{5}, \Gamma_{7},  \Gamma_{9}
& \rightarrow & (\gamma\cdot l^{(3\pm)}),  & \gamadt\pm\sigmats^{(2)}
\end{array}
\end{eqnarray}
where
\begin{eqnarray}
s^{(2\pm)} & = & (0,0,\pm s_{2},0) \nonumber \\
s^{(3\pm)} & = & (0,0,0,\pm s_{3}) \nonumber \\
l^{(2\pm)} & = & (l_{0},0,\pm l_{0},0) \nonumber \\
l^{(3\pm)} & = & (l_{0},0,0,\pm l_{0})
\end{eqnarray}


\section{Appendix - Solutions of the equations}
\label{app:eqnsols}


\subsection{Solutions of the transverse equation}
\label{appsubs:Teqnsols}

We consider plane wave solutions
that satisfy the transverse \schstu equation as well, of the form
\begin{equation}
\phi_{\perp\tau t}(p)\sim e^{-i\frac{p^{2}_{\perp}}{2M}\tau+ipx}u_{\perp}(p,t)
\end{equation}
where $u_{\perp}(p,t)$ is a four-spinor dependent on the specific momentum
and vector $t_{\mu}$.
We observe that Eqs.\ (\ref{eq:phichi}),(\ref{eq:chiphi}) commute with the
operator
\begin{equation}
h(p,t)\equiv\helicity  \label{eq:helicity}
\end{equation}
This operator was shown by Horwitz and Arshansky \cite{HA82} to correspond
to helicity in the frame where $t_{\mu}=(1,0,0,0)$,
\begin{equation}
\helicity\;\;\longrightarrow\;\; {\bf \frac{\Sigma\cdot p}{\mid p\mid}}
\end{equation}
so we shall call $h(p,t)$ the extended helicity operator. An interesting
feature of the
extended helicity operator is that in the reference frame where
$p_{\mu}=(p_{0},0,0,0)$,
we get
\begin{equation}
\helicity\;\;\longrightarrow\;\; {\bf \frac{\Sigma\cdot t}{\mid t\mid}}
\end{equation}
This is because in this frame $\mid p_{\perp}\mid=+p_{0}\sqrt{{\bf t}^{2}}$.
The operator $h(p,t)$ can be decomposed
into two partial operators $h(p,t)=h_{5}(t)\cdot h_{p}(p,t)$ which are
\begin{eqnarray}
h_{5}(t) & \equiv & \hfive \\
h_{p}(p,t) & \equiv & \hpt
\end{eqnarray}
and where they all commute.
The operator $h_{5}(t)$ when looked upon in the frame where
$t_{\mu}=(1,0,0,0)$
is seen to be
\begin{equation}
h_{5}(t) \;\; \longrightarrow \;\; i\gamaz\gamaf =
\left(
\begin{array}{cc}
0 & i \\
-i & 0
\end{array}
\right)
\end{equation}
It exchanges the left and right handed parts of the wave function, and gives
them
a relative phase factor. The left part is rotated counterclockwise by $\pi /2$
in the complex plane and the right part is rotated clockwise by the same
amount.
We call this operator the chiral precedence operator because its eigenvalues
indicate which chiral part precedes the other relative to a counterclockwise
rotation
in the complex plane. This is an interesting feature of the solutions of
the equations of motion
since it expresses an inherent broken symmetry between the left and the right
parts of the wave function. Moreover, it gives physical meaning to the
{\sl relative} phase of the wave function's components, and creates a
coherence
which is preserved through the evolution. This feature of the wave function
is independent of its momentum.

We assumed a four-spinor as the basic structure of the solutions, therefore
we can
use two of these operators to characterize completely the four available
solutions.
The chiral precedence operator (for this purpose the product of
helicity operator and chiral precedence operator has the same effect)
also transforms Eq.\ (\ref{eq:eqtn1}), the equation of motion, into
Eq.\ (\ref{eq:eqtn2}), the equation of motion with interchanged nilpotents,
and vice versa, therefore relating $\psip$ to $\varphip$
\begin{equation}
\hfive\psip=\varphip
\end{equation}
This way, the structure of the main field (and auxiliary field) is set to
conform to the eigenvectors of the chiral precedence projection
operator

\begin{equation}
P_{h_{5}\pm}=\frac{1}{2}({\bf 1}\pm\hfive)
\end{equation}
where $P^{2}_{h_{5}\pm}=P_{h_{5}\pm}$. This projection operator is connected to
the projection part
of one of the nilpotent options discussed in Appendix \ref{app:all_potents}
Eqs.\ (\ref{eq:availnils2}),
and is non-equivalent to the nilpotents we use, since we cannot
transform from one to the other by a unitary transformation.
\par The solutions are seen to be
\begin{eqnarray}
\zeta_{\perp}^{+}(t) \equiv \left( \begin{array}{c}
i\sigma^{\mu} t_{\mu}\xi^{+}  \\
\xi^{+}
\end{array} \right)
&
\zeta_{\perp}^{-}(t) \equiv \left( \begin{array}{c} \xi^{-} \\
i\tilde{\sigma}^{\mu}t_{\mu}\xi^{-}
\end{array} \right)
\label{eq:transols1}
\end{eqnarray}
where the superscript $\pm$ denotes the eigenvalue of the eigenvectors when
operated upon by the chiral precedence operator. When $P_{h_{5}+}$ operates
on  $\zeta_{\perp}^{+}(t)$ it gives $1$, and on $\zeta_{\perp}^{-}(t)$ it
gives $0$. The opposite happens for $P_{h_{5}-}$.
In the special frame where $t_{\mu}=(1,0,0,0)$ the solutions are
\begin{eqnarray}
\zeta_{\perp}^{+} = \left( \begin{array}{c} i\xi^{+}  \\
\xi^{+}
\end{array} \right)
&
\zeta_{\perp}^{-} = \left( \begin{array}{c} \xi^{-} \\
i\xi^{-}
\end{array} \right) \label{eq:sol1}
\end{eqnarray}
and $\xi^{\pm}$ can be taken as two independent vectors, for example
\begin{equation}
\begin{array}{cc}
\xi^{+}=\left( \begin{array}{c}
1 \\
0
\end{array}
\right)
&
\xi^{-}=\left( \begin{array}{c}
0 \\
1
\end{array}
\right)
\end{array}
\end{equation}

Since each spinor can be decomposed and characterized by the helicity
projection operator
\begin{equation}
P_{h\pm}=\frac{1}{2}({\bf 1}\pm\helicity) \label{eq:projhel}
\end{equation}
where $P^{2}_{h\pm}=P_{h\pm}$, and by the chiral precedence operator, we
use them to characterize the four available solutions for a specific momentum.

In order to find the general form of the solutions, we consider the projection
operators $P_{h\pm}$ and $P_{h_{5}\pm}$ which commute. We can form the
following non-normalized solutions
\begin{equation}
u_{\perp}^{rs}(p,t)=\frac{1}{2}({\bf 1}+\epsilon_{s}\helicity)
\zeta_{\perp}^{r}(t) \label{eq:gensolutions}
\end{equation}
where $\zeta_{\perp}^{r}(t)$ are boosted four-spinors composed of stacked
$\xi^{\pm}$ two-spinors as in Eq.\ (\ref{eq:sol1}),
and $\epsilon_{s}$ is $\pm 1$ depending on the eigenvalues of the extended
helicity operator. We define the normalization constant $N(p,t)$ so that
\begin{equation}
N^{2}(p,t)\bar{u}_{\perp}^{r^{\prime}s^{\prime}}(p,t)\gdt
u_{\perp}^{rs}(p,t)=\delta_{rr^{\prime}}\delta_{ss^{\prime}}
\label{eq:normal1}
\end{equation}

We turn now to compute the total integrated probability density, which
should be unity.
\begin{equation}
\int d^{4}x\; \rho_{\perp\tau t}(x)=\int d^{4}x\;
(\bar{\phi}_{\perp\tau t}(x)\gdt\phi_{\perp\tau t}(x)) \label{eq:totprob}
\end{equation}
A wave packet is
\begin{equation}
\phi_{\perp\tau t}(x)=\int d^{4}p\; N(p,t)\sum_{rs}c(r,s,p^{2},p\cdot t)\;
u_{\perp}^{rs}(p,t)e^{-i\frac{p_{\perp}^{2}}{2M}\tau+ipx}
\end{equation}
where $c(r,s,p^{2},p\cdot t)$ are the weights of the wave packet's components.
Performing the $d^{4}x$ integration in
Eq.\ (\ref{eq:totprob}) and one $d^{4}p$ integration, we obtain
\begin{eqnarray}
\int d^{4}x\; \rho_{\perp\tau t}(x) & = & (2\pi)^{4}\int d^{4}p\;N^{2}(p,t)
\sum_{r,r^{\prime},s,s^{\prime}}c^{\ast}(r^{\prime},s^{\prime},p^{2},p\cdot t)
c(r,s,p^{2},p\cdot t) \nonumber \\
&   & \times \bar{u}_{\perp}^{r^{\prime}s^{\prime}}(p,t)\gdt
u_{\perp}^{rs}(p,t)
\end{eqnarray}
Using Eq.\ (\ref{eq:normal1}), we find (one can say that Eq.\
(\ref{eq:Tnormto1}) results in a constant
depending on $t_{\mu}$ rather than $1$,
and the integration over $d^{4}t\;\delta(t^{2}+1)$ is normalized to $1$)
\begin{eqnarray}
\int d^{4}x\; \rho_{\perp\tau t}(x) = (2\pi)^{4}\int d^{4}p\;\sum_{r,s} \mid
c(r,s,p^{2},p\cdot t)\mid^{2} = 1 \label{eq:Tnormto1}
\end{eqnarray}

We compute the total integrated currents on all space-time and all $t_{\mu}$,
from Eq.\ (\ref{eq:fincurrs}), by using
$j_{\perp\tau}^{\mu}=-\frac{i}{2M}(\bar{\phi}_{\perp\tau t}\gdt
\stackrel{\leftrightarrow}{\partial_{\perp}^{\mu}}\phi_{\perp\tau t})$.
Then we have
\begin{eqnarray}
J_{\perp\mu}=\int d^{4}x\;d^{4}t\;\delta(t^{2}+1)\;j_{\perp\tau\mu}(x) & = &
\int d^{4}x\;d^{4}t\;\delta(t^{2}+1)\; \frac{-i}{2M}(\bar{\phi}_{\perp\tau t}
(x)\gdt\LRdmudt\phi_{\perp\tau t}(x)) \nonumber \\
& = & (2\pi)^{4}\int d^{4}p\;d^{4}t\;\delta(t^{2}+1)\frac{p_{\perp\mu}}{M}
\sum_{r,s} \mid c(r,s,p^{2},p\cdot t)\mid^{2} \nonumber \\
\label{eq:totcur}
\end{eqnarray}
so the integrated current is
\begin{equation}
J_{\perp\mu}=\langle \frac{p_{\perp\mu}}{M}\rangle
\end{equation}
Now, $p_{\perp\mu}$ is space-like, and so is $j_{\perp\mu}$.
Since we perform the integration over $t_{\mu}$, thus having various
components of $p_{\mu}$, it is possible
to acquire a time-like $J_{\perp\mu}$ current, as in \cite{HAR90}.

The equations of motion,
Eqs.\ (\ref{eq:phichi}),(\ref{eq:chiphi}), describe the evolution of the
state of a four dimensional
momentum and currents, existing in the plane orthogonal to the time-like
vector
$t_{\mu}$. Considering the frame where $t_{\mu}=(1,0,0,0)$, $p_{\perp\mu}$
becomes
$p_{i}$ and we are dealing with a space-like current.
The appearance of $P_{\mu}$ as a four-vector in the theory is purely
for reasons of keeping the invariance of the theory.
Another manifestation of this aspect is presented in Section
VII, where we
find the Hermitian term $\sigma_{\small{t}}^{\mu\nu}f_{\mu\nu}$, which
transforms to $\sigma^{ij}$ in the frame where $t_{\mu}=(1,0,0,0)$,
and that is why there is no problem of Hermiticity in the second order charged
equation. The momentum of an event can be
thought of as being partitioned by $t_{\mu}$ into a longitudinal part and a
transverse part. Each part evolves according to a different evolution
equation, and they refer to the same vectors $p_{\mu}$, $t_{\mu}$.


\subsection{Solutions of the longitudinal equation}
\label{appsubs:Leqnsols}

For the longitudinal equation we consider the plane wave solutions of the form
\begin{equation}
\phi_{\parallel\tau t}(p)\sim e^{-i\frac{p^{2}_{\parallel}}{2M}\tau+ipx}
u_{\parallel}(p,t)
\end{equation}
Again the equations of motion, Eqs.\ (\ref{eq:lphichi}),(\ref{eq:lchiphi})
commute
with the helicity operator from Eq.\ (\ref{eq:helicity}), so it can be used
to characterize the solutions. Two more operators that commute with the
equations of motion are
\begin{eqnarray}
h_{t}(t) & \equiv & \htt \\
h_{pt5}(p,t) & \equiv & \hptf
\end{eqnarray}
where $h_{pt5}=h_{t}(p,t)\cdot h(p,t)$, and they all commute. Notice that
\begin{eqnarray}
h_{t}(t) & = & i\gamaf h_{5}(t) \nonumber \\
h_{pt5}(p,t) & = & i\gamaf h_{p}(p,t) \nonumber
\end{eqnarray}
\par The characterization is by the projection of helicity Eq.\
(\ref{eq:projhel})
and the projection operator
\begin{equation}
P_{h_{t}\pm}=\frac{1}{2}({\bf 1}\pm\htt)
\end{equation}
The operator $h_{t}(t)$ in the frame where $t_{\mu}=(1,0,0,0)$ is seen to be
just $\gamaz$ which is the parity operator; we denote it as the extended
parity operator. Therefore the characterization
of the solutions of the longitudinal equation is done by extended helicity
and extended parity. The extended parity operator
also transforms Eq.\ (\ref{eq:leqtn1}) into
its counterpart with nilpotents exchanged, and vice versa. We obtain the
relation
\begin{equation}
\htt\psil=\varphil
\end{equation}
The solutions are seen to be
\begin{eqnarray}
\zeta_{\parallel}^{+}(t) \equiv \left( \begin{array}{c}
\sigma^{\mu} t_{\mu}\xi^{+}  \\
\xi^{+}
\end{array} \right)
&
\zeta_{\parallel}^{-}(t) \equiv \left( \begin{array}{c}
\xi^{-} \\
-\tilde{\sigma}^{\mu}t_{\mu}\xi^{-}
\end{array} \right)
\label{eq:longsols1}
\end{eqnarray}
where the superscript $\pm$ denotes the eigenvalue of the eigenvectors when
operated upon by the extended parity operator. In the special frame where
$t_{\mu}=(1,0,0,0)$ the solutions are
\begin{eqnarray}
\zeta_{\parallel}^{+} = \left( \begin{array}{c}
\xi^{+}  \\
\xi^{+}
\end{array}  \right)
&
\zeta_{\parallel}^{-} = \left( \begin{array}{c}
\xi^{-} \\
-\xi^{-}
\end{array} \right) \label{eq:sol2}
\end{eqnarray}
and $\xi^{\pm}$ can be taken as two independent vectors.
The general form of the non-normalized spinor solutions is
\begin{equation}
u_{\parallel}^{rs}(p,t)=\frac{1}{2}({\bf 1}+\epsilon_{s}\helicity)
\zeta_{\parallel}^{r}(t) \label{eq:lgensolutions}
\end{equation}
where $\epsilon_{s}$ has the same meaning as in the transverse case,
and $r$ denotes the eigenvalues of extended parity.

As for the transverse case we set the normalization condition on the
probability density
\begin{equation}
\int d^{4}x\; \rho_{\parallel\tau t}(x)=\int d^{4}x\;
(\bar{\phi}_{\parallel\tau t}(x)\gdt\phi_{\parallel\tau t}(x))
\label{eq:totprob2}
\end{equation}
so that
\begin{equation}
N^{2}(p,t)\bar{u}_{\parallel}^{r^{\prime}s^{\prime}}(p,t)\gdt
u_{\parallel}^{rs}(p,t)=\delta_{rr^{\prime}}\delta_{ss^{\prime}}
\label{eq:normal2}
\end{equation}
A wave packet is
\begin{equation}
\phi_{\parallel\tau t}(x)=\int d^{4}p\; N(p,t)\sum_{r,s}d(r,s,p^{2},p\cdot t)
\; u_{\parallel}^{rs}(p,t)e^{-i\frac{p_{\parallel}^{2}}{2M}\tau+ipx}
\end{equation}
and we get after the integration over $d^{4}x$ and one over $d^{4}p$
\begin{eqnarray}
\int d^{4}x\; \rho_{\parallel\tau t}(x) & = & (2\pi)^{4}\int d^{4}p\;
N^{2}(p,t) \sum_{r,r^{\prime},s,s^{\prime}}d^{\ast}(r^{\prime},s^{\prime},
p^{2},p\cdot t)d(r,s,p^{2},p\cdot t) \nonumber \\
&   & \times \bar{u}_{\parallel}^{r^{\prime}s^{\prime}}(p,t)\gdt
u_{\parallel}^{rs}(p,t)
\end{eqnarray}
Using Eq.\ (\ref{eq:normal2}) we find
\begin{eqnarray}
\int d^{4}x\; \rho_{\parallel\tau t}(x) = (2\pi)^{4}\int d^{4}p\;\sum_{r,s}
\mid d(r,s,p^{2},p\cdot t)\mid^{2} = 1  \label{eq:norma2}
\end{eqnarray}
As for the integrated currents using,
$j_{\parallel\tau}^{\mu}=-\frac{i}{2M}(\phibarl\gdt
\stackrel{\leftrightarrow}{\partial_{\parallel}^{\mu}}\phil)$,
we find
\begin{eqnarray}
J_{\parallel\mu}=\int d^{4}x\;d^{4}t\;\delta(t^{2}+1)\;j_{\parallel\tau\mu}
(x) & = & \int d^{4}x\;d^{4}t\;\delta(t^{2}+1)\;\frac{-i}{2M}
(\bar{\phi}_{\parallel\tau t}(x)\gdt\LRdmudl\phi_{\parallel\tau t}(x))
\nonumber \\
& = & (2\pi)^{4}\int d^{4}p\;d^{4}t\;\delta(t^{2}+1)\frac{p_{\parallel\mu}}
{M}\sum_{r,r^{\prime},s,s^{\prime}}\mid
d(r,s,p^{2},p\cdot t)\mid^{2} \nonumber \\
\label{eq:totcur2}
\end{eqnarray}
so the integration of the current (over $t_{\mu}$ as well) is
\begin{equation}
J_{\parallel\mu}=\langle \frac{p_{\parallel\mu}}{M}\rangle
\end{equation}
In a frame where $t_{\mu}=(1,0,0,0)$, the zeroth component of the currents
$j_{\parallel\mu}$
is the only one which survives due to the vanishing of $\partial_{\perp i}$
in this frame, so the currents
are time-like. In accordance with Lorentz invariance this is true in any
reference frame.
As in the transverse case, $J_{\parallel\mu}$ may be time-like, depending on
the relative weights of the wave packet's components.


\section{Appendix - Discrete symmetries}
\label{app:pct}

When dealing with the discrete symmetries of the theory it is necessary to
know how the pre-Maxwell field $a_{\alpha}$, $(\alpha=0,1,2,3,\tau)$
transforms. We require that the free kinetic term of the pre-Maxwell field
(see \cite{HAE88}),
$\frac{1}{4}\lambda f^{\alpha\beta}f_{\alpha\beta}$, be invariant.
Furthermore, we define some new forms of solutions
for both the transverse and longitudinal equations, which are just the ones
defined in Appendix \ref{app:eqnsols} but transformed, and are helpful for
a full discussion.
The basic form of the solutions of the transverse equations are given in
Eq.\ (\ref{eq:transols1}); we redefine them by choosing $\xi^{\pm}$ to be
an eigenvector of $-i\sigma^{2}$ with eigenvalues of $\pm i$
\begin{equation}
\begin{array}{cc}
\xi^{+}=\left( \begin{array}{c}
i \\
1
\end{array}
\right)
&
\xi^{-}=\left( \begin{array}{c}
1 \\
i
\end{array}
\right)
\end{array}
\end{equation}
Multiplying each component of $\zeta_{\perp}^{+}(t)$
by $-i\tilde{\sigma}^{\mu}t_{\mu}$, and each component of
$\zeta_{\perp}^{-}(t)$
by $-i\sigma^{\mu}t_{\mu}$, and exchanging $\xi^{+}$ by $\xi^{-}$, we get
\begin{eqnarray}
\eta_{\perp}^{+}(t) \equiv \left( \begin{array}{c} \xi^{-} \\
-i\tilde{\sigma}^{\mu}t_{\mu}\xi^{-}
\end{array} \right)
&
\eta_{\perp}^{-}(t) \equiv \left( \begin{array}{c}
-i\sigma^{\mu} t_{\mu}\xi^{+}  \\
\xi^{+}
\end{array} \right)
\label{eq:transols2}
\end{eqnarray}
The basic form of the solutions of the longitudinal equations is given in
Eq.\ (\ref{eq:longsols1}), by multiplying
both components of $\zeta_{\parallel}^{+}(t)$ by
$\tilde{\sigma}^{\mu}t_{\mu}$,
and $\zeta_{\parallel}^{-}(t)$ by $-\sigma^{\mu}t_{\mu}$,
and exchanging $\xi^{+}$ by $\xi^{-}$, we get
\begin{eqnarray}
\eta_{\parallel}^{+}(t) \equiv \left( \begin{array}{c}
\xi^{-} \\
\tilde{\sigma}^{\mu}t_{\mu}\xi^{-}
\end{array} \right)
&
\eta_{\parallel}^{-}(t) \equiv \left( \begin{array}{c}
-\sigma^{\mu} t_{\mu}\xi^{+}  \\
\xi^{+}
\end{array} \right)
\label{eq:longsols2}
\end{eqnarray}
We now proceed to the treatment of parity, charge conjugation, and $\tau$
reversal.

\subsection{Parity}
\label{subapp:parity}

The transverse charged equations of motion exhibit a symmetry under the
inversion of parity. The transformation is
\begin{eqnarray}
{\bf x} & \rightarrow & -{\bf x} \nonumber \\
{\bf P} & \rightarrow & -{\bf P} \nonumber \\
{\bf t} & \rightarrow & -{\bf t} \nonumber \\
{\bf a}(x) & \rightarrow & {\bf a}(x^{\prime}) = -{\bf a}(x)
\end{eqnarray}
Applying it to the transverse equations of motion we get
\begin{eqnarray}
(-\sigma^{0i}\Pi_{0}t_{i}-\sigma^{i0}\Pi_{i}t_{0}+\sigma^{ij}\Pi_{i}t_{j})
\phi_{\perp\tau t_{0},-{\bf t}}^{\prime}(x^{\prime}) & = & \splus M(\gamaz
t_{0}-{\bf \gamma}\cdot{\bf t})\chi_{\perp\tau t_{0},-{\bf t}}^{\prime}
(x^{\prime}) \nonumber \\
(-\sigma^{0i}\Pi_{0}t_{i}-\sigma^{i0}\Pi_{i}t_{0}+\sigma^{ij}\Pi_{i}t_{j})
\chi_{\perp\tau t_{0},-{\bf t}}^{\prime}(x^{\prime}) & = & \sminus(\gamaz
t_{0}-{\bf\gamma}\cdot{\bf t})(\idtau+e\atau)\phi_{\perp\tau t_{0},
-{\bf t}}^{\prime}(x^{\prime}) \nonumber \\
\end{eqnarray}
Multiplying by $\gamaz$ from the left we get back the original form of
equations
\begin{eqnarray}
\sigpit\gamaz\phi_{\perp\tau t_{0},-{\bf t}}^{\prime}(x^{\prime}) &
= & \splus M\gdt\gamaz\chi_{\perp\tau t_{0},-{\bf t}}^{\prime}(x^{\prime})
\nonumber \\
\sigpit\gamaz\chi_{\perp\tau t_{0},-{\bf t}}^{\prime}(x^{\prime}) &
= & \sminus\gdt(\idtau+e\atau)\gamaz\phi_{\perp\tau t_{0},-{\bf t}}^{\prime}
(x^{\prime})
\end{eqnarray}
and the wave functions transform (up to a phase) as
\begin{eqnarray}
\phi^{P}_{\perp\tau t}(x)=\gamaz\phi_{\perp\tau t_{0},-{\bf t}}(x^{0},
-{\bf x}) \nonumber \\
\chi^{P}_{\perp\tau t}(x)=\gamaz\chi_{\perp\tau t_{0},-{\bf t}}(x^{0},
-{\bf x})
\end{eqnarray}
where the superscript $P$ denotes parity.
The same is true for the charged longitudinal equations of motion, where
we find the wave functions transforming as
\begin{eqnarray}
\phi^{P}_{\parallel\tau t}(x)=\gamaz\phi_{\parallel\tau t_{0},-{\bf t}}(x^{0},
-{\bf x}) \nonumber \\
\chi^{P}_{\parallel\tau t}(x)=\gamaz\chi_{\parallel\tau t_{0},-{\bf t}}(x^{0},
-{\bf x})
\end{eqnarray}
The solutions of the transverse equations are constructed from a helicity
projection operator, and a $\zeta_{\perp}^{\pm}(t)$ part.
A similar situation exists for the longitudinal equations.
We check both components.

When we check the action of parity on the $\zeta_{\perp}^{\pm}(t)$ parts of
the solutions of the transverse equation, we find
\begin{eqnarray}
P\zeta_{\perp}^{+}(t)\equiv\gamaz\zeta_{\perp}^{+}(t_{0},-{\bf t}) & = &
\zeta_{\perp}^{-}(t) \nonumber \\
P\zeta_{\perp}^{-}(t)\equiv\gamaz\zeta_{\perp}^{-}(t_{0},-{\bf t}) & = &
\zeta_{\perp}^{+}(t)
\end{eqnarray}
and
\begin{eqnarray}
P\eta_{\perp}^{+}(t)\equiv\gamaz\eta_{\perp}^{+}(t_{0},-{\bf t}) & = &
\eta_{\perp}^{-}(t) \nonumber \\
P\eta_{\perp}^{-}(t)\equiv\gamaz\eta_{\perp}^{-}(t_{0},-{\bf t}) & = &
\eta_{\perp}^{+}(t)
\end{eqnarray}
On the other hand,
for the solutions of the longitudinal equations, we find
\begin{eqnarray}
P\zeta_{\parallel}^{+}(t)\equiv\gamaz\zeta_{\parallel}^{+}(t_{0},-{\bf t}) &
= & \eta_{\parallel}^{+}(t) \nonumber \\
P\zeta_{\parallel}^{-}(t)\equiv\gamaz\zeta_{\parallel}^{-}(t_{0},-{\bf t}) &
= & \eta_{\parallel}^{-}(t)
\end{eqnarray}
and
\begin{eqnarray}
P\eta_{\parallel}^{+}(t)\equiv\gamaz\eta_{\parallel}^{+}(t_{0},-{\bf t}) & =
& \zeta_{\parallel}^{+}(t) \nonumber \\
P\eta_{\parallel}^{-}(t)\equiv\gamaz\eta_{\parallel}^{-}(t_{0},-{\bf t}) & =
& \zeta_{\parallel}^{-}(t)
\end{eqnarray}
We can make a combination of states for which this transformation makes them
transfrom one into the other. This combination is useful also for charge
conjugation and $\tau$ reversal. The combination is
\begin{eqnarray}
P(\zeta_{\parallel}^{+}(t) \pm \eta_{\parallel}^{-}(t)) & = &
(\zeta_{\parallel}^{-}(t) \pm \eta_{\parallel}^{+}(t)) \nonumber \\
P(\zeta_{\parallel}^{-}(t) \pm \eta_{\parallel}^{+}(t)) & = &
(\zeta_{\parallel}^{+}(t) \pm \eta_{\parallel}^{-}(t))
\end{eqnarray}
Doing the same thing for the transverse case one obtaines
\begin{eqnarray}
P(\zeta_{\perp}^{+}(t) \pm \eta_{\perp}^{-}(t)) & = & (\zeta_{\perp}^{-}(t)
\pm \eta_{\perp}^{+}(t)) \nonumber \\
P(\zeta_{\perp}^{-}(t) \pm \eta_{\perp}^{+}(t)) & = & (\zeta_{\perp}^{+}(t)
\pm \eta_{\perp}^{-}(t))
\end{eqnarray}
On the other hand, when we check the action of normal parity on the helicity
projection operator part of the solutions, we find
\begin{equation}
\gamaz P_{h\pm}(p_{0},-{\bf p},t_{0},-{\bf t})=P_{h\mp}(p,t)\gamaz
\end{equation}
therefore the helicity flips sign.

Another type of parity is what we call generalized parity, denoted by
${\cal P}$,
including the
regular parity and a time inversion. Usually the discrete symmetry
under time inversion is treated separately and differently from that of
parity,
as an anti-hermitian operator. In our theory time is just another dimension
distinguished from the space dimensions by the metric tensor, and the
evolution
is governed by $\tau$. Therefore the role of time as described by Wigner is
transferred to $\tau$, the time symmetry becoming much simpler. This
generalized parity transformation is
\begin{eqnarray}
x_{\mu} & \rightarrow & -x_{\mu} \nonumber \\
P_{\mu} & \rightarrow & -P_{\mu} \nonumber \\
t_{\mu} & \rightarrow & -t_{\mu} \nonumber \\
a_{\mu} & \rightarrow & -a_{\mu} \label{eq:genparity}
\end{eqnarray}
When we come to define the way the wave functions transform we have a few
options. The charged form of Eqs.\ (\ref{eq:phichi}),(\ref{eq:chiphi})
\begin{eqnarray}
\gdt\sigpit\phip & = & \splus M\chip \nonumber \\
\gdt\sigpit\chip & = & 2\sminus(\idtau+e\atau)\phip \label{eq:fullch}
\end{eqnarray}
permits two
compensating operators for the transformation of Eqs.\ (\ref{eq:genparity}),
$\gamaf$, and $\gdt$. We prefer to use $\gamaf$ as a generalized parity
operator because it is not coupled to any four-vector.
Applying the transformation we find
\begin{eqnarray}
-\gdt\sigpit\phi_{\perp\tau,-t}^{\prime}(x^{\prime}) & = &
\splus M\chi_{\perp\tau,-t}^{\prime}(x^{\prime}) \nonumber \\
-\gdt\sigpit\chi_{\perp\tau,-t}^{\prime}(x^{\prime}) & = &
2\sminus(\idtau+e\atau)\phi_{\perp\tau,-t}^{\prime}(x^{\prime})
\end{eqnarray}
and after multiplying by $\gamaf$ from the left we obtain the usual form
\begin{eqnarray}
\gdt\sigpit\gamaf\phi_{\perp\tau,-t}^{\prime}(x^{\prime}) & = &
\splus M\gamaf\chi_{\perp\tau,-t}^{\prime}(x^{\prime}) \nonumber \\
\gdt\sigpit\gamaf\chi_{\perp\tau,-t}^{\prime}(x^{\prime}) & = &
2\sminus(\idtau+e\atau)\gamaf\phi_{\perp\tau,-t}^{\prime}(x^{\prime})
\end{eqnarray}
In this case, the wave functions transform (up to a phase) as
\begin{eqnarray}
\phi^{{\cal P}}_{\perp\tau t}(x)=\gamaf\phi_{\perp\tau ,-t}(-x) \nonumber \\
\chi^{{\cal P}}_{\perp\tau t}(x)=\gamaf\chi_{\perp\tau ,-t}(-x)
\end{eqnarray}
A similar transformation holds for the longitudinal equations of motion.
As can be seen by operating on the solutions of transverse
and longitudinal equations, we get
\begin{eqnarray}
{\cal P}\zeta_{\perp,\parallel}^{+}(t)\equiv\gamaf
\zeta_{\perp,\parallel}^{+}(-t) & = & +\zeta_{\perp,\parallel}^{+}(t)
\nonumber \\
{\cal P}\zeta_{\perp,\parallel}^{-}(t)\equiv\gamaf
\zeta_{\perp,\parallel}^{-}(-t) & = & -\zeta_{\perp,\parallel}^{-}(t)
\end{eqnarray}
and
\begin{eqnarray}
{\cal P}\eta_{\perp,\parallel}^{+}(t)\equiv\gamaf
\eta_{\perp,\parallel}^{+}(-t) & = & -\eta_{\perp,\parallel}^{+}(t)
\nonumber \\
{\cal P}\eta_{\perp,\parallel}^{-}(t)\equiv\gamaf
\eta_{\perp,\parallel}^{-}(-t) & = & +\eta_{\perp,\parallel}^{-}(t)
\end{eqnarray}
Therefore, the generalized parity operation on the transverse equations,
brings the set of solutions into themselves.

Here too, we can use combinations which behave under generalized parity
the same as shown for the regular parity, but now the behavior of the
transverse and longitudinal solutions is the same. For the transverse
and longitudinal cases we have
\begin{eqnarray}
{\cal P}(\zeta_{\perp,\parallel}^{+}(t) \pm \eta_{\perp,\parallel}^{-}(t)) &
= & +(\zeta_{\perp,\parallel}^{+}(t) \pm \eta_{\perp,\parallel}^{-}(t))
\nonumber \\
{\cal P}(\zeta_{\perp,\parallel}^{-}(t) \pm \eta_{\perp,\parallel}^{+}(t)) &
= & -(\zeta_{\perp,\parallel}^{-}(t) \pm \eta_{\perp,\parallel}^{+}(t))
\label{eq:gpt}
\end{eqnarray}
Note that the combinations
\begin{eqnarray}
\zeta^{+}(t)+\eta^{-}(t) \nonumber \\
\zeta^{-}(t)-\eta^{+}(t)
\end{eqnarray}
are pure right handed, and the combinations
\begin{eqnarray}
\zeta^{+}(t)-\eta^{-}(t) \nonumber \\
\zeta^{-}(t)+\eta^{+}(t)
\end{eqnarray}
are pure left handed, for both $\perp$ and $\parallel$.
The extended helicity operator
remains unchanged under ${\cal P}$, so there is no helicity flip in this case.


\subsection{Charge conjugation}
\label{subapp:chargeconj}

The theory is symmetric under charge conjugation. We define
the charge conjugation operation, denoted by ${\cal C}$,
as the transformation necessary
to bring the charged equations of motion to the same form only with the sign
of $e$ reversed. First conjugate the equations
\begin{eqnarray}
\gdt^{\ast}(\sigma^{\mu\nu\ast}(-P_{\mu}-ea_{\mu})t_{\nu})
\phi_{\perp\tau t}^{\prime\ast}(x^{\prime}) & = & \splus M
\chi_{\perp\tau t}^{\prime\ast}(x^{\prime}) \nonumber \\
\gdt^{\ast}(\sigma^{\mu\nu\ast}(-P_{\mu}-ea_{\mu})t_{\nu})
\chi_{\perp\tau t}^{\prime\ast}(x^{\prime}) & = &
2\sminus(-\idtau+e\atau)\phi_{\perp\tau t}^{\prime\ast}(x^{\prime})
\end{eqnarray}
then make the substitution
\begin{eqnarray}
\atau & \rightarrow & -\atau \nonumber \\
\tau  & \rightarrow & -\tau
\end{eqnarray}
which is required to bring the equations to the correct form. This is
consistent
with the invariance of the free kinetic term of the pre-Maxwell field
$\frac{1}{4}\lambda f^{\alpha\beta}f_{\alpha\beta}$.
The next step is to multiply them by $(i\gamaf\gamatwo)=(\gamaf C\gamaz)$
where $C=i\gamatwo\gamaz$, because
\begin{eqnarray}
i\gamaf\gamatwo(\gamamu^{\ast})(i\gamaf\gamatwo)^{-1} & = & \gamamu
\nonumber \\
i\gamaf\gamatwo(\sigmunu^{\ast})(i\gamaf\gamatwo)^{-1} & = & -\sigmunu
\end{eqnarray}
One then obtains
\begin{eqnarray}
\gdt\tsigpit i\gamaf\gamatwo\phi_{\perp-\tau t}^{\prime\ast}(x^{\prime}) & =
& \splus M i\gamaf\gamatwo\chi_{\perp-\tau t}^{\prime\ast}(x^{\prime})
\nonumber \\
\gdt\tsigpit i\gamaf\gamatwo\chi_{\perp-\tau t}^{\prime\ast}(x^{\prime}) & =
& 2\sminus(\idtau-e\atau)i\gamaf\gamatwo\phi_{\perp-\tau t}^{\prime\ast}
(x^{\prime})
\end{eqnarray}
where $\tilde{\Pi}_{\mu}=P_{\mu}+ea_{\mu}$.
Taking $a_{\alpha}\rightarrow -a_{\alpha}$ brings us back to the original
form. Therefore, the wave functions transform (up to a phase) as
\begin{eqnarray}
\phi^{{\cal C}}_{\perp\tau t}(x) & = & i\gamaf\gamatwo
\phi{^\ast}_{\perp,-\tau,t}(x) \nonumber \\
\chi^{{\cal C}}_{\perp\tau t}(x) & = & i\gamaf\gamatwo
\chi^{\ast}_{\perp,-\tau,t}(x)
\end{eqnarray}
The same persists for the longitudinal equations. This transformation includes
a $\tau$ reversal for its consistency.
When the pre-Maxwell
field radiation equations \cite{SHA89} are taken into consideration,
\begin{eqnarray}
-\dtau f^{\tau\mu}+\dnu f^{\mu\nu} & = & ej^{\mu} \nonumber \\
\dmu f^{\tau\mu}=ej^{\tau} & = & e\rho
\end{eqnarray}
a symmetry breakdown is observed in the second equation due the need to
change the sign of the $\atau$ field. This is a consequence of the
Schr\"{o}dinger type equation used.
Here $\rho$ is a scalar. One could think of a Klein-Gordon generalization
for which $j_{\tau}\propto\psi^{\ast}\!\!\!\stackrel{\leftrightarrow}
{\partial_{\tau}}\!\!\!\psi$
for which this problem would not occur (note that the same question arises
in the non-relativistic Schr\"{o}dinger quantum theory in interaction with
the standard Maxwell field).

To see the consequences of performing charge conjugation to the solutions of
the transverse equation, we define solutions as in
Eq.\ (\ref{eq:gensolutions}), using $\eta_{\perp}^{r}(t)$,
\begin{equation}
w_{\perp}^{rs}(p,t)=\frac{1}{2}({\bf 1}+\epsilon_{s}\helicity)
\eta_{\perp}^{r}(t) \label{eq:wsols}
\end{equation}
The subscript $s$ in Eq.\ (\ref{eq:wsols}) denotes the eigenvalue
of the extended helicity operator. Considering a free event, we have for the
helicity projection operator
\begin{equation}
i\gamaf\gamatwo P_{h\pm}^{\ast}=P_{h\mp}i\gamaf\gamatwo
\end{equation}
and since
\begin{eqnarray}
-i\sigma^{2}\xi^{+\ast}=-\xi^{-} \nonumber \\
-i\sigma^{2}\xi^{-\ast}=+\xi^{+}
\end{eqnarray}
for the $\zeta_{\perp}^{r}(t)$ part of the solutions, we have
\begin{eqnarray}
{\cal C}\zeta_{\perp}^{+}(t) & = & -\eta_{\perp}^{+}(t) \nonumber \\
{\cal C}\zeta_{\perp}^{-}(t) & = & +\eta_{\perp}^{-}(t)
\end{eqnarray}
and
\begin{eqnarray}
{\cal C}\eta_{\perp}^{+}(t) & = & +\zeta_{\perp}^{+}(t) \nonumber \\
{\cal C}\eta_{\perp}^{-}(t) & = & -\zeta_{\perp}^{-}(t)
\end{eqnarray}
so we can use the combination
\begin{eqnarray}
{\cal C}(\zeta_{\perp}^{+}(t) \pm \eta_{\perp}^{-}(t)) & = &
-(\zeta_{\perp}^{-}(t) \pm \eta_{\perp}^{+}(t))  \nonumber \\
{\cal C}(\zeta_{\perp}^{-}(t) \pm \eta_{\perp}^{+}(t)) & = &
+(\zeta_{\perp}^{+}(t) \pm \eta_{\perp}^{-}(t))
\end{eqnarray}
These combinations have mixed eigenvalues of the chiral precedence operator,
positive for $\zeta_{\perp}^{+}(t)$, $\eta_{\perp}^{+}(t)$, and negative for
$\zeta_{\perp}^{-}(t)$, $\eta_{\perp}^{-}(t)$. Therefore for the
combination we obtain
\begin{equation}
{\cal C}e^{-i\frac{p^{2}_{\perp}}{2M}\tau+ipx}(u_{\perp}^{r,s}(p,t)\pm
w_{\perp}^{-r,s}(p,t))=-\epsilon_{r}e^{-i\frac{p^{2}_{\perp}}{2M}\tau-ipx}
(w_{\perp}^{r,-s}(p,t)\pm u_{\perp}^{-r,-s}(p,t))
\end{equation}
where $\epsilon_{r}$ is $\pm 1$ depending on value of $r$, either $+$ or $-$.
The state transforms into the other helicity state. Since no space inversion
was performed, we get a helicity flip. Therefore, the charge conjugated
wave function
describes an event with opposite charge and spin, moving in the opposite
direction.
As mentioned in Appendix \ref{subapp:parity} dealing with parity, these
combinations
are pure left or right handed spinors. For example, the charge conjugated
left handed spinor is a right handed spinor with opposite charge moving in the
opposite direction.

Repeating these operations for the solutions of the longitudinal equations, we
find similar results, we have only to change the subscript $\perp$ to
$\parallel$.


\subsection{$\tau$ reversal}
\label{subapp:taurev}

Here we discuss $\tau$ reversal in the sense introduced by Wigner; reversing
the evolution parameter we reverse the motion of the event, thus creating
a current in the opposite direction.
The operations needed for the $\tau$ reversal symmetry, denoted as
${\cal T}$, result
in the same operations as those for the charge conjugation as can be seen by
performing the operation
\begin{equation}
\tau\rightarrow -\tau
\end{equation}
on the equations of motion. To compensate we have to take
\begin{equation}
a_{\mu}\rightarrow -a_{\mu}
\end{equation}
and applying this to Eqs.\ (\ref{eq:fullch}) while taking the conjugate of the
equations,  we get the transformation properties of the wave functions
\begin{eqnarray}
\phi^{{\cal T}}_{\perp\tau t}(x) & = & \gamaf\gamatwo
\phi{^\ast}_{\perp,-\tau,t}(x) \nonumber \\
\chi^{{\cal T}}_{\perp\tau t}(x) & = & \gamaf\gamatwo
\chi^{\ast}_{\perp,-\tau,t}(x)
\end{eqnarray}
As usual, the longitudinal equations have the same transformation properties.
We can repeat the same arguments given for the charge conjugation.
Although the transformation looks the same as the charge conjugation one,
the reason for doing it stems from Wigner's idea of time reversal. However,
we obtain a charge conjugated event.


\subsection{${\cal PCT}$ invariance}
\label{subapp:pctinvar}

The most striking thing about these discrete transformations is the identity
of the charge conjugation transformation, and $\tau$ reversal transformation
in the sense of Wigner. Conceptually
this is understood using Feynman diagrams. We know from Dirac's theory that
a charge conjugated electron, i.e. a positron, moves backward in space-time
with opposite charge. This is exactly what a $\tau$ reversal means. Seen in
the
four dimensional world (a projection of the trajectory on space-time at a
specific $\tau$), a $\tau$ reversal changes the charge, direction, and
helicity. Concluding that charge conjugation and $\tau$ reversal have the
same
consequences, a full ${\cal PCT}$ transformation is just a generalized parity
transformation. We get an event with opposite motion in space-time
relative to the state before conjugation, with the
same helicity and charge. The symmetry under generalized parity is a direct
consequence of the manifest covariance built into the theory.

\section{Appendix - Dirac matrices}
\label{app:diracmat}

Our orthogonal matrix basis which satisfies the Dirac Clifford algebra is
\cite{Schweb,Thal}
\begin{equation}
\label{eq:gamamat}
\begin{array}{llll}
\Gamma_{1}={\bf 1} & \Gamma_{2}=\gamaz & \Gamma_{3}=i\gamaone & \Gamma_{4}=
i\gamatwo \nonumber  \\
\Gamma_{5}=i\gamatre & \Gamma_{6}=\gamaz\gamaone & \Gamma_{7}=\gamaz\gamatwo
& \Gamma_{8}=\gamaz\gamatre  \nonumber \\
\Gamma_{9}=i\gamatwo\gamatre & \Gamma_{10}=i\gamatre\gamaone & \Gamma_{11}=
i\gamaone\gamatwo & \Gamma_{12}=\gamaone\gamatwo\gamatre \nonumber \\
\Gamma_{13}=i\gamaz\gamatwo\gamatre & \Gamma_{14}=i\gamaz\gamatre\gamaone &
\Gamma_{15}=i\gamaz\gamaone\gamatwo & \Gamma_{16}=i\gamaz\gamaone\gamatwo
\gamatre
\end{array}
\end{equation}
We use the following chiral representation of the Dirac matrices
\begin{eqnarray}
\gamaz = \left(  \begin{array}{cc}
0 & 1 \\
1 & 0
\end{array}
\right)
&
\gamma^{i} = \left( \begin{array}{rc}
\!\!0 & \sigma^{i} \\
-\sigma^{i} & 0
\end{array}
\right)
\end{eqnarray}
and
\begin{eqnarray}
\gamaf = \left( \begin{array}{rc}
-1 & 0 \\
0 & 1
\end{array}
\right)
\end{eqnarray}


\end{document}